\documentclass[fleqn,usenatbib]{mnras}

\usepackage{macros}
\usepackage{subfig}
\usepackage{graphicx}
\usepackage{threeparttable}
\usepackage{tcolorbox}
\usepackage[ruled,vlined]{algorithm2e}
\usepackage{amsmath,amssymb}
\usepackage{lineno}
\usepackage{hyperref}
\graphicspath{{./}{./figures/}}
\newcommand{\edit}[1]{{\textcolor{black}{#1}}}
\newcommand{\refresponse}[1]{{\textcolor{black}{#1}}}

\title[PSF Higher Moments Error and WL II]{Impact of Point Spread Function Higher Moments Error on Weak Gravitational Lensing II: A Comprehensive Study}

\author[]{Tianqing Zhang$^1$, Husni Almoubayyed$^1$, Rachel Mandelbaum$^1$,  Joshua~E.~Meyers$^2$, \newauthor Mike Jarvis$^3$,  Arun Kannawadi$^4$, Morgan~A. Schmitz$^5$, Axel Guinot$^6$, \newauthor The LSST Dark Energy Science Collaboration.
 \\
$^1$ McWilliams Center for Cosmology, Department of Physics, Carnegie Mellon University, 5000 Forbes Ave, Pittsburgh, PA 15213, USA.\\
$^2$ Lawrence Livermore National Laboratory, Livermore, CA 94551. \\
$^3$ Department of Physics and Astronomy, University of Pennsylvania, Philadelphia, PA 19104, USA.\\
$^4$ Department of Astrophysical Sciences, Princeton University, 4 Ivy Lane, Princeton, NJ 08544, USA.\\
$^5$ Universit\'e C\^{o}te d’Azur, Observatoire de la C\^{o}te d’Azur, CNRS, Laboratoire Lagrange, Bd de l’Observatoire, CS 34229, 06304 Nice \\Cedex 4,France.\\
$^6$ Université de Paris, CNRS, Astroparticule et Cosmologie, F-75013 Paris, France.
}

\date{\today}

\let\oldequation\equation
\let\oldendequation\endequation

\renewenvironment{equation}
  {\linenomathNonumbers\oldequation}
  {\oldendequation\endlinenomath}

\let\oldalign\align
\let\oldendalign\endalign

\renewenvironment{align}
  {\linenomathNonumbers\oldalign}
  {\oldendalign\endlinenomath}

\begin{document}

\label{firstpage}
\pagerange{\pageref{firstpage}--\pageref{LastPage}}
\maketitle

\begin{abstract}
Weak lensing is one of the most powerful probes for dark matter and dark energy science, although it faces increasing challenges in controlling systematic uncertainties as the statistical errors become smaller. The Point Spread Function (PSF) needs to be precisely modeled to avoid systematic error on the weak lensing measurements. The weak lensing biases induced by errors in the PSF model second moments, i.e., its size and shape, are well-studied. However, \citet{2021arXiv210705644Z} showed that errors in the higher moments of the PSF may also be a significant source of systematics for upcoming weak lensing surveys. Therefore, this work comprehensively investigate the modeling quality of PSF moments from the $3^{\text{rd}}$ to $6^{\text{th}}$ order, and propagate the \textsc{PSFEx} higher moments modeling error in the HSC survey dataset to the weak lensing shear-shear correlation functions and their cosmological analyses. The overall multiplicative shear bias associated with errors in PSF higher moments can cause a $\sim 0.1 \sigma$ shift on the cosmological parameters for LSST Y10, while the associated additive biases can induce $1\sigma$ uncertainties in cosmology parameter inference for LSST Y10, if not accounted. We compare the \textsc{PSFEx} model with PSF in Full FOV (\textsc{Piff}), and find similar performance in modeling the PSF higher moments. We conclude that PSF higher moment errors of the future PSF models should be reduced from those in current methods, otherwise needed to be explicitly modeled in the weak lensing analysis.
\end{abstract}


\section{Introduction}
\label{sec:introduction}

Weak gravitational lensing, or weak lensing, is the slight deflection of the light from distant objects by the gravitational effect of nearby objects. Weak lensing leads to a mild change in the object's shape, size and flux, and it is a powerful probe of the dark matter distribution of the Universe due to its sensitivity to the gravitational potential along the line of sight \citep{Hu:2001fb, 2010GReGr..42.2177H, 2013PhR...530...87W}. To date, the most promising way of measuring weak lensing is to measure its coherent effects on the galaxy shape, i.e., the weak lensing shear.  Weak lensing can be caused by a nearby massive galaxy or cluster, i.e.,  as measured using galaxy-galaxy lensing \citep[e.g.,][]{2014MNRAS.437.2111V, 2015MNRAS.454.1161Z, 2018PhRvD..98d2005P}; or by the large-scale  structure of the Universe, as measured using cosmic shear \citep[e.g.,][]{2020PASJ...72...16H, 2021A&A...645A.104A, 2021arXiv210513543A}. 

The coherent galaxy shape distortions caused by weak lensing are \edit{currently} measured using millions, \edit{in the future even billions,} of galaxies in large astronomical surveys. The ``Stage III'' cosmological surveys \citep{Albrecht:2006um} that started in the previous decade provided weak lensing observation that moved the field forward substantially; these include the Dark Energy Survey \citep[DES;][]{des_review}, the Kilo-Degree Survey \citep[KiDS;][]{deJong:2017bkf}, and the Hyper Suprime-Cam survey \citep[HSC;][]{2018PASJ...70S...4A}. In the near future, ``Stage IV'' surveys will begin to observe at greater depth and/or area than the previous generation; the Stage IV surveys include the Vera C. Rubin Observatory Legacy Survey of Space and Time \citep[LSST;][]{Ivezic:2008fe, 2009arXiv0912.0201L}, the \textit{Nancy Grace Roman} Space Telescope High Latitude Imaging Survey \citep{2015arXiv150303757S, 2019arXiv190205569A} and \textit{Euclid} \citep{Euclid_overview}. These new surveys will provide greater statistical precision in the measurements, and therefore demand greater control of systematic uncertainties in weak lensing.

The Point Spread Function (PSF) is the function that describes the \edit{atmospheric turbulence}, telescope optics, and some detector effects \citep{2000PASP..112.1360A, 2013A&A...551A.119P} \edit{on a point source image}.  PSF modeling algorithms reconstruct the PSF \edit{at the position of the} stars, and interpolate the model to arbitrary positions \edit{on the image, e.g., \textsc{PSFEx} \citep{2011ASPC..442..435B}, or to positions on the sky, e.g., \textsc{Piff} (PSF in Full FOV; \citealt{2021MNRAS.501.1282J})}.

The raw light profile of the galaxies \edit{is convolved with the PSF}, changing their observed shapes and sizes. Since measuring weak lensing signals relies heavily on measuring the coherent galaxy shape distortions, modeling the PSF correctly is fundamental for controlling weak lensing systematics. Failure of the PSF model to represent the true PSF causes systematic errors in the inferred galaxy shapes and weak lensing shears. Previous studies have developed a formalism that cleanly describes how the errors in modeling PSF second moments, i.e., the shape and size, affect the galaxy shape measurement and weak lensing shear inference \citep[e.g.,][]{Hirata:2003cv, PaulinHenriksson:2007mw, 2010MNRAS.404..350R,2016MNRAS.460.2245J}. There is also a formalism that describes how the PSF second moment errors further propagate to the weak lensing observables (shear-shear correlations), using the ``$\rho$-statistics'' \citep{2010MNRAS.404..350R, 2016MNRAS.460.2245J}. 

However, the aforementioned formalism, which is commonly used for quantifying the quality of PSF modeling, does not consider the impact on weak lensing shear caused by errors in the higher moments, i.e., moments with order higher than the second, of the PSF model. In \citet{2020A&A...636A..78S}, excess multiplicative and additive shear bias is found in addition to the predictions of the second moment formalism, for \textit{Euclid}'s PSF. 
A previous study by \citet{2021arXiv210705644Z} (\edit{hereafter} ZM21) explored this topic by carrying out shape measurement experiments, with the radial kurtosis of the PSF intentionally mis-modeled, while preserving the PSF second moments. \edit{They} found that errors in the PSF radial kurtosis can induce a multiplicative bias in the inferred weak lensing shear. \edit{They also} found that for parametric galaxy models based on the COSMOS survey, and for PSF radial kurtosis errors as in the HSC public data release 1 \citep[PDR1;][]{2018PASJ...70S...8A} PSF models from \textsc{PSFEx} \citep{2011ASPC..442..435B}, the PSF radial kurtosis error can cause a redshift-dependent multiplicative shear bias at the level of the LSST Y10 requirement \citep{2018arXiv180901669T},  
thus motivating further research on this topic. 

In this paper, we want to extend the understanding from ZM21 in several ways:  (a) include a wider range of PSF higher moments, which might induce both multiplicative and additive shear biases; (b) propagate the biases into the common weak lensing data vector, the two-point correlation function (2PCF) \edit{$\xi_{\pm}$}, and to cosmological parameter estimates; (c) include \textsc{Piff}, which might provide some estimate of how algorithm-dependent the errors in PSF higher moments are, and might serve as a better example of an algorithm that will be used for LSST.

\edit{We introduce background material, including the  weak lensing shear, PSF higher moments, and shapelet decomposition in Section~\ref{sec:background}.
In Section~\ref{sec:data}, we describe the HSC datasets in this work for measuring the PSF higher moments, and show the results of the PSF modeling quality on the second and higher moments for two PSF models, \textsc{PSFEx} and \textsc{Piff}. In Section~\ref{sec:simulation}, we describe the methodology of single galaxy simulations, including simulation workflow, galaxy and PSF profiles, and how we change the PSF higher moments with the aid of shapelet decomposition. We also show the results based on these single galaxy simulations. In Section~\ref{sec:analyses}, we combine the results from Section~\ref{sec:data} and \ref{sec:simulation} to further propagate the systematics induced by PSF higher moment errors to the weak lensing 2PCF, and its associated cosmology analyses by Fisher forecasting. In Section~\ref{sec:conclusions}, we discuss the implications of our results for weak lensing with future imaging surveys. }

\section{Background}
\label{sec:background}

In this section, we describe the background of this paper. \edit{In Section~\ref{sec:background:wl}, we introduce the formalism to quantify the weak lensing shear.}  In Section~\ref{sec:background:mom_measure_mtd}, we introduce the method for measuring the higher moments of PSFs. We then introduce the radial shapelet decomposition, used as a basis in which we expand any given PSF light profile, in Section~\ref{sec:background:shapelet}. 

\subsection{Weak Lensing}
\label{sec:background:wl}

Weak gravitational lensing, or weak lensing, is the coherent gravitational distortion on background (source) galaxy flux, size, and shape by foreground (lens) objects. The lens can be any massive object, e.g., a galaxy cluster, or the cosmic large-scale structure.  
Weak lensing is a powerful observable because of its sensitivity to the matter distribution along the line of sight \citep{Hu:2001fb,2010GReGr..42.2177H,2013PhR...530...87W}. 
In this paper, we are interested in the cosmic shear, which is the coherent distortion of the source galaxy shapes by the large-scale structure of the Universe, resulting in a nonzero two-point correlation function of galaxy shapes. The distortion of the galaxies by the weak lensing shear is \edit{determined by the reduced shear $g  = g_1 + i g_2$, which is a combination of the shear and the convergence \citep{Mandelbaum:2017jpr}}.
$g_1$ describes the shear along the x- or y-axes, 
while $g_2$ describes the shear along \edit{an angle $\pi/4$ defined by growing counterclockwise from} the x-axis \edit{on the image}. Here the x-y axes are aligned with the local (RA, Dec) axes on the sky. 

For a cosmological weak lensing analysis, it is useful to measure the weak lensing two-point correlation function \citep{1991ApJ...380....1M}, also referred to as the 2PCF. We can calculate the shear along a chosen angular vector $\boldsymbol{\theta}$  
connecting two galaxies, with polar angle $\phi$, by $g_t = -\mathcal{R}(g e^{-2i\phi})$, and $\pi/4$ to $\boldsymbol{\theta}$ by $g_\times = -\mathcal{I}(g e^{-2i\phi})$. The shear 2PCF is computed by
\begin{equation}
\label{eq:2pcf_def}
\xi_\pm (\theta) = \langle g_t g_t \rangle (\theta) \pm \langle g_\times g_\times \rangle (\theta).
\end{equation}
Since the weak lensing shear is isotropic (statistically speaking), the $\xi_\pm (\theta)$ is integrated over the polar angle \edit{$\phi$} and presented as a function of the angular distance $\theta=|\boldsymbol{\theta}|$. 

The weak lensing shear 2PCF as measured \edit{through} $\xi_\pm$ is sensitive to the coherent change in galaxy shapes due to large-scale structure  \citep{2002A&A...396....1S}, though it is contaminated by intrinsic alignments  \citep[e.g.,][]{2000ApJ...545..561C, 2000MNRAS.319..649H, 2015PhR...558....1T, 2015SSRv..193....1J}, i.e., the correlated galaxy alignments due to local effects such as tidal fields. 

Estimating shear accurately is a key step in any cosmological analysis of weak lensing data. Shear biases are commonly  
modeled as two terms, the multiplicative bias $m$ and the additive bias $c$ \citep{2006MNRAS.368.1323H,Massey:2006ha}, which enter the estimated shear as
\begin{equation}
\label{eq:shear_biases_def}
    \hat{g} = (1+m)g + c,
\end{equation}
where $\hat{g}$ denotes the estimated shear. Systematic biases in the estimated shear must not exceed a certain portion of the statistical error to avoid substantial biases in the reported constraints on the cosmological parameters compared to those that would ideally be recovered. 
\edit{We are particularly interested in a redshift-dependent multiplicative bias; as suggested in \citet{2013MNRAS.429..661M}, 
a redshift-dependent multiplicative bias can bias the inferred dark energy equation of state parameter from weak lensing. This is motivated since ZM21 found that the shear response to the PSF higher moment errors depends on the galaxy properties, which means that the galaxy ensemble in each tomographic bin will respond differently to the same PSF higher moment error.  
In \citet{2018arXiv180901669T}, the redshift-dependent multiplicative bias is parameterized by $m_0$ in 
\begin{equation}\label{eq:m0}
    m(z) = m_0 \left(\frac{2z - z_\text{max}}{z_\text{max}}\right) + \bar{m},
\end{equation}
\edit{where $\bar{m}$ is a non-zero average multiplicative bias over redshift.} Error budget requirements are placed on the upper bound of the absolute value of multiplicative biases for weak lensing surveys \citep{2016MNRAS.460.2245J, 2018PASJ...70S..25M}.  
Taking LSST Y10 as an example \citep{2018arXiv180901669T}, the requirement on the redshift-dependent multiplicative bias, which is the difference in $m$ across the full source redshift range, is 0.003. This motivates detailed studies on the connection between weak lensing shear systematics and other factors, including the PSF higher-moment modeling error (ZM21 and this work).} \edit{Note that we only discuss the PSF-induced multiplicative shear biases in this work, without other sources of redshift-dependent multiplicative biases \citep[e.g.,][]{2022MNRAS.509.3371M}.}

\subsection{Moment Measurement}
\label{sec:background:mom_measure_mtd}

In this section, we introduce the methods for measuring higher moments of the PSF. Firstly, we define the adaptive second moment $\mathbf{M}$ for a light profile,
\begin{equation}
\label{eq:second_moment}
    M_{pq} = \frac{\int \mathrm{d}x \, \mathrm{d}y \, x^p \, y^q \, \omega(x,y) \, I(x,y)}{\int \mathrm{d}x \, \mathrm{d}y \, \omega(x,y) \, I(x,y) },
\end{equation}
\edit{where $(p,q) = (2,0)$, $(1,1)$, or $(0,2)$.} Here $I(x,y)$ is the image intensity, where $\mathbf{x} = (x,y)$ is the image coordinate with origin at the centroid of $I(x,y)$. $\omega(x,y)$ in Eq.~\eqref{eq:second_moment} is the adaptive Gaussian weight, which has the same second moments as the light profile $I(x,y)$ \citep{Hirata:2003cv}, defined by 
\begin{equation}
\label{eq:def_weight_function}
\omega(\mathbf{x}) = \text{exp}[-\mathbf{x}^T \mathbf{M}^{-1} \mathbf{x}].
\end{equation}

The second moment size $\sigma$ and shape $e_1$ and $e_2$ can then be calculated from the second moments $\mathbf{M}$ using
\begin{align}
\label{eq:def_sigma}
    \sigma &= \left[\text{det}(\mathbf{M})\right]^{\frac{1}{4}}\\\label{eq:def_e1}
    e_1 &= \frac{M_{20} - M_{02}}{M_{20}+ M_{02}}\\\label{eq:def_e2}
    e_2 &= \frac{2 M_{11}}{M_{20} + M_{02}}.
\end{align}
Here $\text{det}(\mathbf{M}) = M_{02} M_{20} - M_{11}^2$ is the determinant of the second moment matrix. From Eqs.~\eqref{eq:def_sigma}--\eqref{eq:def_e2}, we can solve for the weighted second moments $M_{ij}$ given the weighted shape $(e_1, e_2)$ and size $\sigma$, which are measured using the  \textsc{HSM} module\footnote{\url{https://galsim-developers.github.io/GalSim/_build/html/hsm.html}} \citep{Hirata:2003cv,Mandelbaum:2005wv} in \textsc{GalSim} \citep{Rowe:2014cza}. 

Based on the second moments, we also define a standardized coordinate system $(u,v)$ in Eq.~\eqref{eq:standard_coord}; this is the coordinate system where the profile $I(u,v)$ has zero second moment shape $e_1=e_2=0$, defined in Eqs.~\eqref{eq:def_e1}--\eqref{eq:def_e2}, 
and second moment size $\sigma = 1$, defined in Eq.~\eqref{eq:def_sigma}.
The standardized coordinate system can be determined via a linear transformation of the image coordinate system as follows:
\begin{equation}
\begin{pmatrix}
u\\
v
\end{pmatrix} = \mathbf{M}^{-\frac{1}{2}} \begin{pmatrix}
x\\
y
\end{pmatrix}=
\begin{pmatrix}
M_{20} & M_{11}  \\
M_{11} & M_{02}
\end{pmatrix}^{-\frac{1}{2}}
\begin{pmatrix}
x\\
y
\end{pmatrix}.
\label{eq:standard_coord}
\end{equation}

The standardized adaptive higher moment, $M_{pq}$, is then defined by
\begin{equation}
\label{eq:moment_define}
    M_{pq} = \frac{\int \mathrm{d}x \, \mathrm{d}y \, [u(x,y)]^p \, [v(x,y)]^q \, \omega(x,y) \, I(x,y)}{\int \mathrm{d}x \, \mathrm{d}y \, \omega(x,y) \, I(x,y) }.
\end{equation}
For the $n$\textsuperscript{th} moments, $p$ takes any value between $0$ to $n$, and $q=n-p$. \edit{We choose to measure PSF higher moments in the standardized coordinate system $(u,v)$ instead of $(x,y)$, as such quantities are scale and shape independent, assuming the PSF is well-sampled. The weight $\omega$ is applied to suppress image noise at large radii during the measurement process. The denominator is the normalizing factor, such that the higher moments will not depend on the amplitudes of the weight and the image.}

\begin{figure}
    \centering
    \includegraphics[width=0.98\columnwidth]{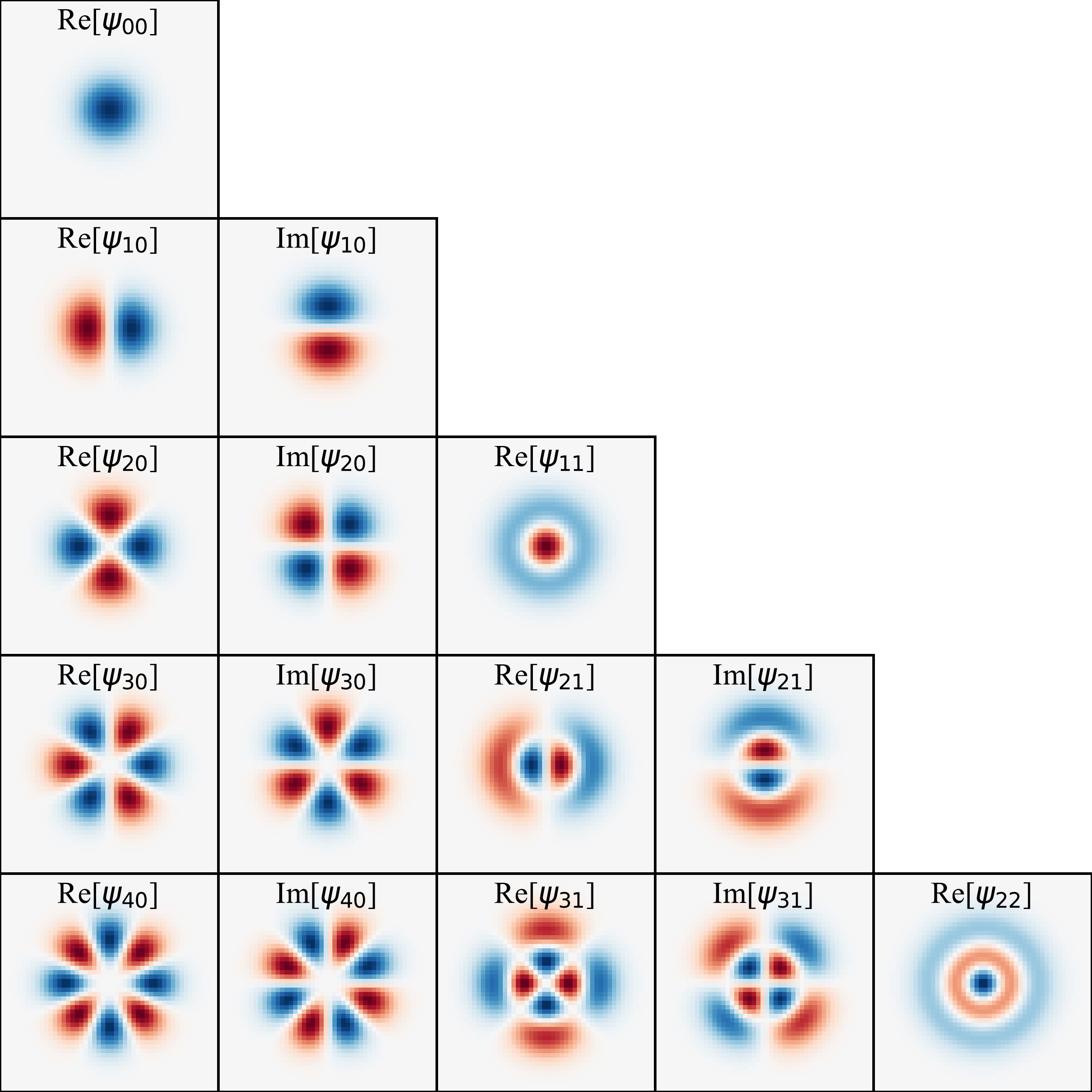}
    \caption{The first 15 unique real and imaginary parts of the shapelet basis functions in Eq.~\eqref{eq:def_shapelet_basis}. 
    We plot the first 5 orders of this basis, i.e., $p+q=0$ through $4$. The color scale for each base covers $[-A,A]$, where $A$ is the maximum of the absolute value of that basis function. }
    \label{fig:laguerre_basis}
\end{figure}

Throughout this paper, we define the biases on the moment $M_{pq}$ as
\begin{equation}
\label{eq:bias_define}
    B[M_{pq}] = M_{pq,\text{model}} - M_{pq,\text{true}},
\end{equation}
where $M_{pq,\text{model}}$ is the moment of the model PSF, and $M_{pq,\text{true}}$ is the moment of the true PSF. Note that we refer to the standardized higher moments as the ``higher moments'' throughout this paper.

\subsection{Shapelet Decomposition}
\label{sec:background:shapelet}

The shapelet decomposition is an expansion of a two-dimensional image with the eigenfunctions of the 2D quantum harmonic oscillator as the basis functions. This basis function is also referred to as the Laguerre Function with Gaussian weight. This method \edit{was} used to expand the galaxy and PSF profile in \citet{2005MNRAS.363..197M} and used to measure weak lensing shear in \citet{2007MNRAS.380..229M}. For detailed explanations of shapelet expansions, see also \citet{Bernstein:2001nz}. In this study, we use the shapelet decomposition implemented in \textsc{GalSim}\footnote{\url{https://github.com/GalSim-developers/GalSim}} \citep{Rowe:2014cza}.

The shapelets basis functions are parameterized by a single parameter: the length scale $L$. After determining the value of $L$ for the image, the image can be decomposed into a series of shapelet coefficients $b_{jk}$, indexed by $j$ and $k$. We also defined two more indices, i.e., the order  $N = j+k$ and the spin number $m = j-k$. The PSF image $I(r,\theta)$ can be expanded by the basis functions of the shapelet coefficients $b_{jk}$, 
\begin{equation}
\label{eq:shapelet_expansion}
I(r, \theta) = \frac{1}{L ^ 2} \sum_{jk} b_{jk} \, \psi_{jk}\left(\frac{r}{L}, \theta\right),
\end{equation}
where $\psi_{jk}\left(\frac{r}{L}, \theta\right)$ is the Laguerre Function with Gaussian weight, i.e., the radial shapelet basis \edit{in a polar coordinate system with radius $r$ and polar angle $\theta$}, 
\begin{equation}
\label{eq:def_shapelet_basis}
    \psi_{jk}(r,\theta) = \frac{-1^q}{\sqrt{\pi}} \sqrt{\frac{j!}{k!}} \, r^m \, e^{im\theta} \, e^{-\frac{r^2}{2}} \, \mathbf{L}_j^{(m)}(r^2).
\end{equation}
The $\mathbf{L}_k^{(m)}(r^2)$ is the Laguerre Polynomial. Fig.~\ref{fig:laguerre_basis} shows the first 15 basis images of $\psi_{pq}$ that we used to decompose the PSF. 
For a given order $N$, there are $2N+1$ shapelet basis functions. Due to conjugate pairings, $N$ of the shapelet coefficients $b_{jk}$ are identical to $b_{kj}$. Therefore, to expand a real image, we have $N+1$ distinct shapelet basis functions for order $N$ \edit{that satisfy $j\geq k$}.

To determine the length scale $L$, we carried out the following experiment: We decomposed the PSF with different length scales $L$; kept the 40 leading $b_{jk}$s of the shapelet series; reconstructed the image using the first forty $b_{jk}$; and measured the residual of this reconstruction. We \edit{found} that to minimize the absolute value of the residual 
of the reconstruction, the length scale $L$ should be set to the weighted second moment $\sigma$ of the PSF defined in Eq.~\eqref{eq:def_sigma}. \edit{This rule was found to be true on both the Gaussian and Kolmogorov profiles.} We therefore adopted this approach throughout this work.


\section{Data}
\label{sec:data}

\begin{figure*}
    \centering
    \includegraphics[width=1.8\columnwidth]{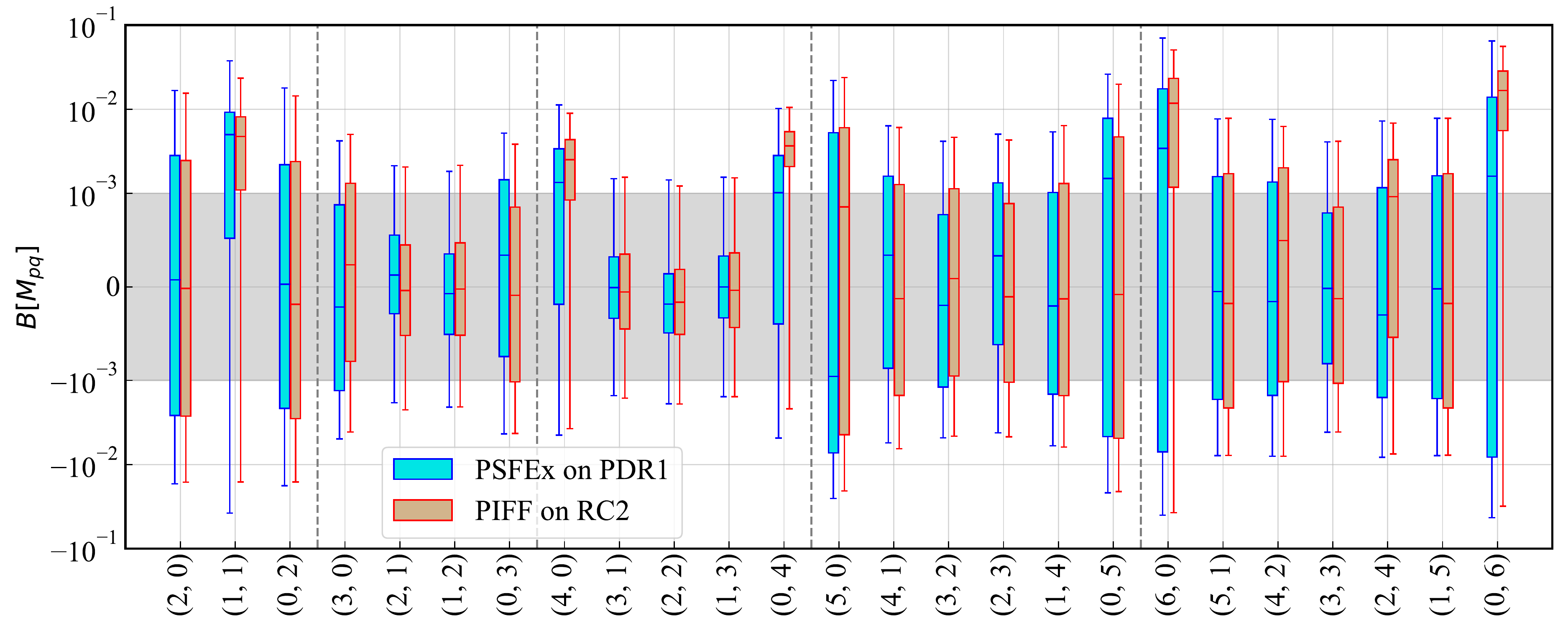}
    \caption{Box plot showing the PSF moment biases from the $2^{\text{nd}}$ to the $6^{\text{th}}$ moments, with the whiskers showing the $2\sigma$ range (from 3rd to 97th percentile), the boxes showing the interquartile range, and the bars showing the median. The \textsc{PSFEx} and \textsc{Piff} results are shown side-by-side. The y-axis is symmetrical log-scaled, with the linear region shown in grey. Although \textsc{PSFEx} and \textsc{Piff} were used to model two different HSC datasets, we observe a comparable order of magnitude in PSF model residuals for the two methods. However, \textsc{Piff}'s median residuals on $M_{40}$, $M_{04}$, $M_{60}$ and $M_{06}$ are a few times larger than those of \textsc{PSFEx}. These are the main contributing higher moments to the shear biases, thus motivating further development of \textsc{Piff}.  
    }
    \label{fig:box_plots}
\end{figure*}

In this section, we introduce the data from the Hyper Suprime-Cam survey \citep[HSC;][]{2018PASJ...70S...4A} to study how well current PSF models recover PSF higher moments. We inspected two datasets, one for \textsc{PSFEx} and one for \textsc{Piff}. 
For both datasets, we used the coadded images of bright stars as the true effective PSF, and compared them with the PSF model at the bright stars' positions. The \textsc{PSFEx} and \textsc{Piff} star catalogs are described in Sections~\ref{sec:data:psfex} and~\ref{sec:data:piff}, respectively. \edit{ We describe the measurement results of the PSF higher moments error in Section~\ref{sec:data:hsc_measure}. }

\subsection{\textsc{PSFEx} Dataset}
\label{sec:data:psfex}

The dataset for \refresponse{quantifying the modeling quality of} \textsc{PSFEx} is the star catalog of the first HSC public data release \citep[PDR1;][]{2018PASJ...70S...8A}. The \textsc{PSFEx} model in this study was generated by the HSC pipeline \citep{2018PASJ...70S...5B} with a modified version of \textsc{PSFEx} \citep{2011ASPC..442..435B}; see Section~3.3 of \citet{2018PASJ...70S...5B} for more details. We used all six fields in the PDR1 survey to inspect the PSF higher moments, instead of just the \code{GAMA\_15H} field as in ZM21.
Our star selection process for the \textsc{PSFEx} is detailed in Section~3.4.1 in ZM21, so we only summarize it briefly here. 

We adopted the ``basic flag cuts'' from Table~3 of \citet{2018PASJ...70S..25M}, with \code{iclassification\_extendedness} set to 0 to identify non-extended objects. These flag cuts eliminate objects that are contaminated or affected by exposure edges, bad pixels, saturation or cosmic rays, and reduce the number of selected stars to $1.1 \times 10^7$.  
We adopted a signal-to-noise ratio (SNR) cut $\text{SNR} > 1000$ to reduce noise in the PSF higher moments measurement, which further reduced the sample size to $3.1 \times 10^5$. \edit{The SNR cut was determined so that the statistical uncertainty in the PSF radial fourth moments of the star images is $<0.1\%$ (ZM21), avoiding a scenario where the higher moments are dominated by the image noise. The i-band magnitudes of the selected stars are between 18 to 20, a regime in which the correction for the brighter-fatter effect \citep{2018PASJ...70S...5B} is highly effective as shown in Section~4.2 of \citet{2018PASJ...70S..25M}.} \refresponse{The SNR selections are only done for our PSF modeling inspection, not when running the PSF modeling step. }

ZM21 identified the need for a cut \code{iblendedness\_abs\_flux}$ > 0.001$ to address the fact that the moment measurements of blended objects are biased. 
In this work, that cut reduced the sample size to $2.6 \times 10^5$. Finally, we also excluded stars with a close neighbor within $2$ arcmin of their centroids using a k-d tree. 
At the end of the selection process, we had $2.4 \times 10^5$ stars, around four times the amount in ZM21 since we used all six HSC fields. The number density of the \textsc{PSFEx} star dataset is $0.62$ arcmin$^{-2}$. Examples of moment residual maps for \textsc{PSFEx} are shown in Appendix~\ref{ap:moment_example}.

\subsection{\textsc{Piff} Dataset}
\label{sec:data:piff}

We measured the performance of \textsc{Piff} \citep{2021MNRAS.501.1282J} on the HSC data in order to compare with \textsc{PSFEx}. \textsc{Piff} was used as the PSF modeling algorithm for the DES Y3 dataset  
and performed  better than previous DES PSF models, especially at modeling continuous trends across multiple detectors.  
\textsc{Piff} has been run on the HSC Release Candidate 2 (RC2)\footnote{Detailed description of the RC2 dataset can be found in \url{https://dmtn-091.lsst.io/v/DM-15448/}.}, which consists of two HSC SSP-Wide tracts and one HSC SSP-UltraDeep tract. 
\refresponse{We used version 1.1.0 of \textsc{Piff}. It modeled PSFs in the image coordinate system, instead of in the WCS coordinates, with pixel scale equal to the native pixel scale ($0.168$~arcsec). The model kernel size is $21\times 21$ pixels. The PSF was interpolated with a second order polynomial. We used $\chi^2$ outlier rejection with \code{nsigma}$ = 4.0$ and \code{max\_remove}$=0.05$. We refer the readers to \cite{2021MNRAS.501.1282J} for a detailed explanation of these settings.  }
The RC2 dataset is reprocessed biweekly using the latest version of Rubin's LSST science pipelines \citep{2017ASPC..512..279J}. We inspected the PSF modeling quality on the two wide-field tracts, 
which correspond to \edit{an area of} $\sim 6$~deg$^2$ (each tract of the HSC data is roughly $3$~\deg$^2$). 

The star selection differs from that used for the \textsc{PSFEx} dataset: we used the \edit{pre-selected} \edit{\textsc{Piff}} candidate stars with SNR$>1000$, without the need for the blending flux and close-neighbor cut. By this criterion, we had in total 11366 stars and PSF models to compare. The number density of the \textsc{Piff} dataset is $0.55$ arcmin$^{-1}$, about 13 per cent lower than that for \textsc{PSFEx}. Examples of moment residual maps for \textsc{Piff} are shown in Appendix~\ref{ap:moment_example}.

\subsection{Measuring PSF Higher Moment Error}
\label{sec:data:hsc_measure}

\begin{figure}
    \centering
    \includegraphics[width=1.0\columnwidth]{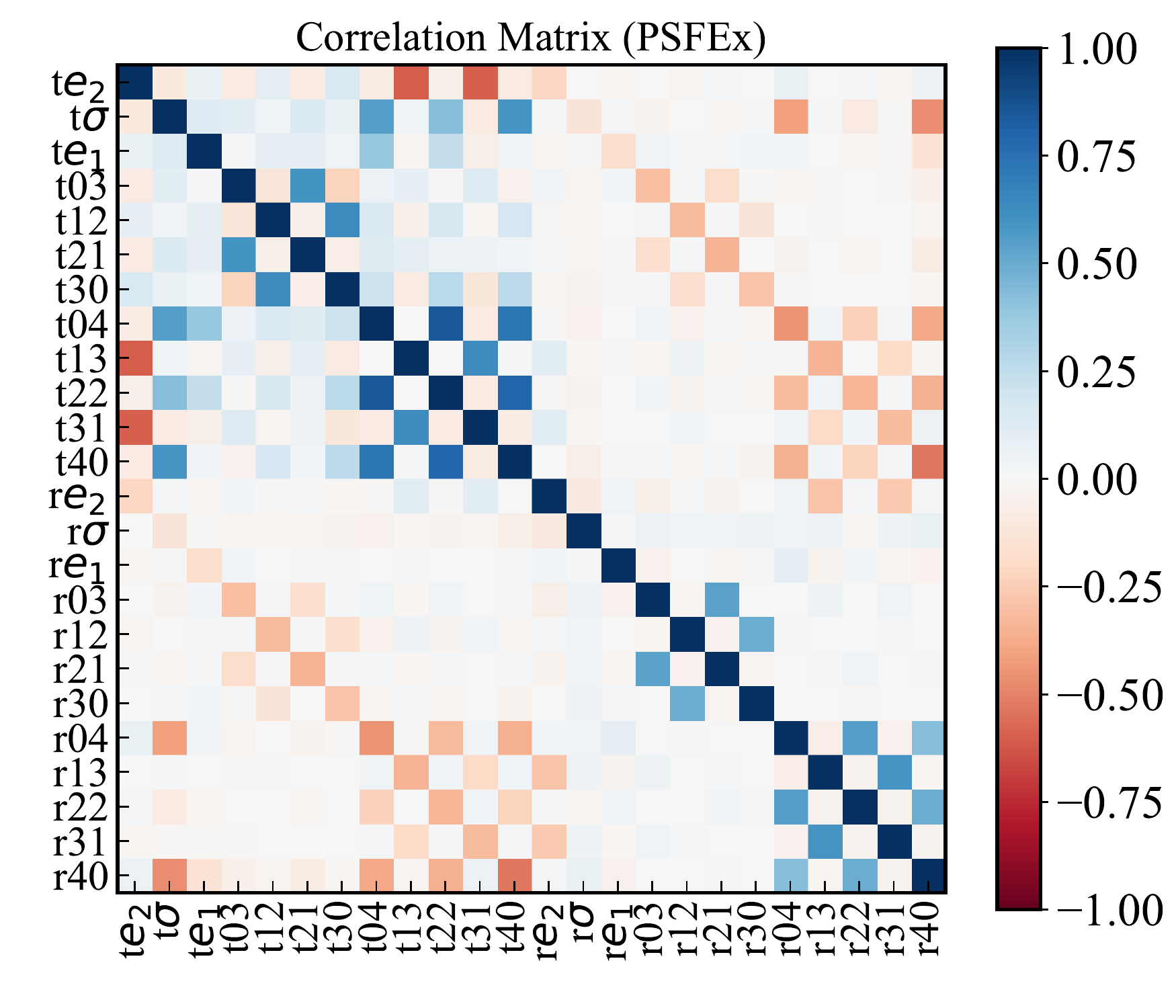}
    \includegraphics[width=1.0\columnwidth]{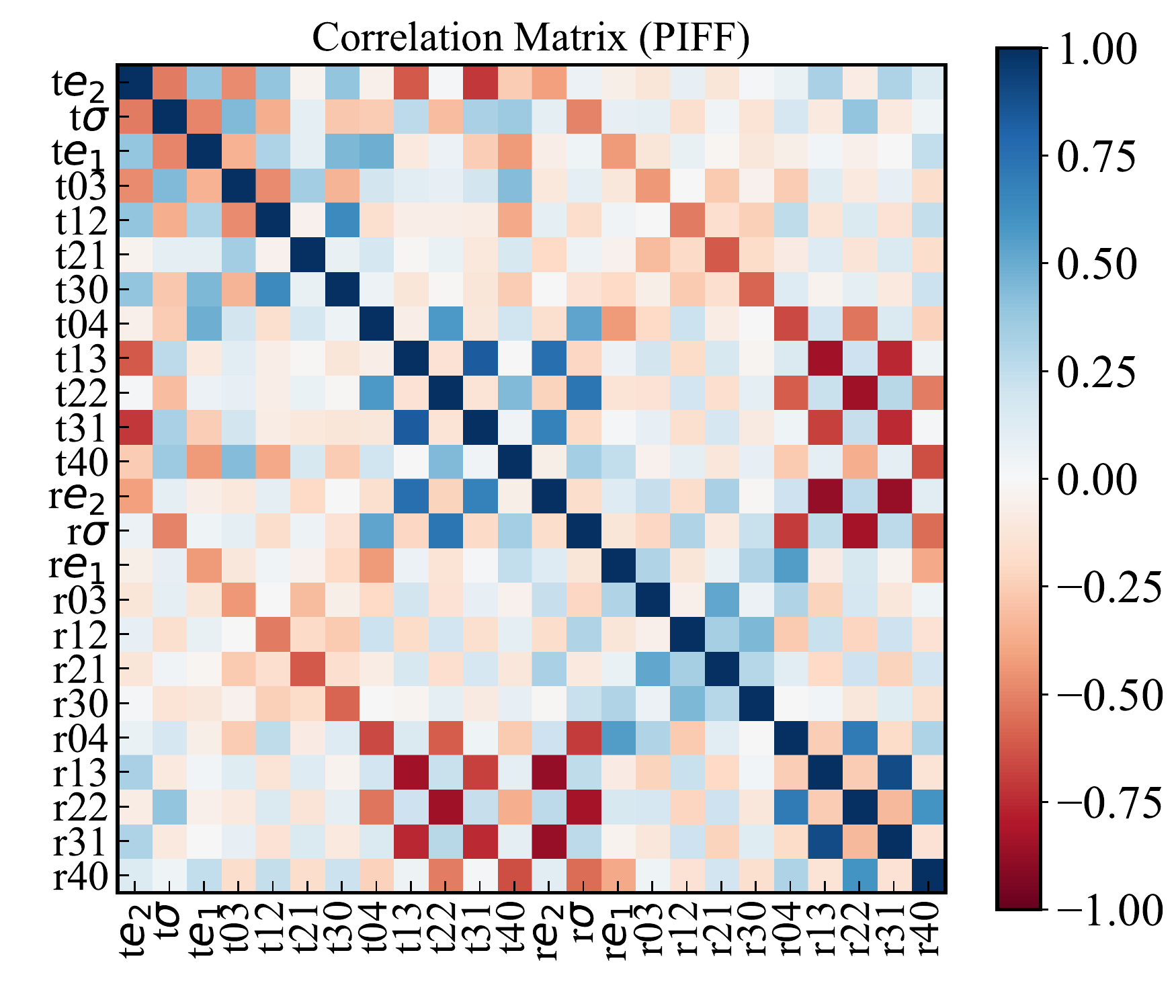}
    \caption{The correlation matrix of \textsc{PSFEx} (upper) and \textsc{Piff} (lower) moments from the 2\textsuperscript{nd} to the 4\textsuperscript{th} moments, where ``t'' denotes  the true values of the moments and ``r'' denotes the moment residuals.}
    \label{fig:correlation_matrix}
\end{figure}

We used the postage stamp images of the selected stars as measures of the true PSF. We obtained the PSF models evaluated at the position of the stars, as the model PSF. We used coadded star images, for which the PSF models are a weighted coaddition of the PSF model in each exposure \citep{2018PASJ...70S...5B}.  
We measured the 22 higher moments, defined in Eq.~\eqref{eq:moment_define}, from the $3^{\text{rd}}$ to the $6^{\text{th}}$ order with the method described in Section~\ref{sec:background:mom_measure_mtd}.  We also measured the weighted second moments with the \textsc{HSM} \citep{Mandelbaum:2005wv} module of \textsc{GalSim}.

We measured the moment biases $B[M_{pq}]$ by subtracting the \edit{star} PSF moments $M_{pq,\text{true}}$ from the model PSF moments $M_{pq, \text{model}}$, as in Eq.~\eqref{eq:bias_define}. In Fig.~\ref{fig:box_plots}, combining the measurements of the PSF higher moments for all of the selected stars in these datasets, we show the distributions of the \textsc{PSFEx} and \textsc{Piff} moment errors $B[M_{pq}]$ with box plots side by side. 
The whiskers of the plot show the $2\sigma$ ranges of the distributions, the boxes show the interquartile ranges, and the bars show the median. We can see from the box plots that the two PSF models have similar PSF second moment residuals, and the PSF sizes are positively biased in both models, as observed for \textsc{PSFEx} in \citet{2018PASJ...70S..25M}.  \refresponse{We listed the mean of the moment residual $\langle B[M_{pq}] \rangle$ for \textsc{PSFEx} and \textsc{PIFF} in Table~\ref{tab:mul_add_single}.}

We calculated the ``bias fluctuation'' field $\widetilde{B}[M_{pq}](\mathbf{x})$ by
\begin{equation}
\label{eq:bias_fluctuation}
\widetilde{B}[M_{pq}](\mathbf{x}) = B[M_{pq}](\mathbf{x}) - \langle B[M_{pq}](\mathbf{x}) \rangle.
\end{equation}
We then used the two-point correlation function (2PCF) to measure the cross-correlation of the bias fluctuations 
$\widetilde{B}[M_{pq}](\mathbf{x})$ and $\widetilde{B}[M_{uv}](\mathbf{x})$,
\begin{equation}
\label{eq:moment_2pcf}
\xi^{pq,uv}(\theta) = \langle \widetilde{B}[M_{pq}](\mathbf{x}) \widetilde{B}[M_{uv}](\mathbf{x} + \theta)  \rangle.
\end{equation}
When $p=u$ and $q=v$, Eq.~\eqref{eq:moment_2pcf} becomes the auto-correlation function of $\widetilde{B}[M_{pq}](\mathbf{x})$. We measured the 2PCFs of the PSF higher moment errors using \textsc{TreeCorr}\footnote{\url{https://github.com/rmjarvis/TreeCorr}} \citep{2004MNRAS.352..338J}.

Because of the relatively small area of the \textsc{Piff} dataset, we only measured  its one-point statistics (mean, covariance matrix, etc.), not its two-point statistics. Therefore, we can only compare \textsc{Piff} with \textsc{PSFEx} at the early analysis stage, rather than propagating to the weak lensing data vector contamination and biases in cosmological parameter estimates. 

\edit{The version of} \textsc{Piff} \edit{used for this work} produces similar order-of-magnitude PSF moment residuals as \textsc{PSFEx} from the 2\textsuperscript{nd} to the 6\textsuperscript{th} moments. However, its median residuals on $M_{40}$, $M_{04}$, $M_{60}$ and $M_{06}$ are several times larger than those for \textsc{PSFEx}, which is important because those are the primary moments contributing to the shear bias. This finding is not surprising because the implementation of \textsc{Piff} integrated with Rubin's LSST Science Pipelines has not been thoroughly tuned, and in particular, none of its testing has focused on its optimization for accurate recovery of PSF higher moments.  
The results for \textsc{Piff} in Fig.~\ref{fig:box_plots} motivate further algorithm development and tuning, by providing additional metrics toward which to optimize in addition to the 2\textsuperscript{nd} moments. \edit{In Appendix~\ref{ap:moment_example:rc2}, we show an apples-to-apples comparison between \textsc{Piff} and \textsc{PSFEx} on the RC2 dataset; the results further motivate the optimization of \textsc{Piff} toward minimizing PSF higher moment residuals. }

In Fig.~\ref{fig:correlation_matrix}, we show the correlation matrix between the true PSF moments and their residuals for \textsc{PSFEx} (upper) and \textsc{Piff} (lower panel). We see a chequered-flag pattern in the correlation matrices. The true  moments with the same parity 
for both $p$ and $q$ are usually positively correlated, and likewise for the residuals. 
This results in a chequered pattern within the same order $n = p+q$ -- the ($p,q$) and the ($p\pm 2, q \mp 2$) moments are correlated -- as well as a bigger chequered pattern across the orders -- between $n$ and $n\pm 2$ orders, though the latter cannot be seen in our plots, since we are only showing $n=3$ and $n=4$ moments. There is an even larger scale pattern: the true moments and residuals for a given $(p,q)$ are typically anti-correlated with each other due to the impact of noise on the true moments. We also observe a significant anti-correlation between ``t$\sigma$'' and ``r04''/ ``r40'' for \textsc{PSFEx}. This indicates that $M_{04}$ and $M_{40}$ are preferentially overestimated in areas of the survey with good seeing.  
This result is consistent with the findings of ZM21, but it is not seen in the \textsc{Piff} results \edit{because it does not perform oversampling for good-seeing images}. However, the correlation matrix of \textsc{Piff} shows stronger anti-correlations between the true and the residual moments, which suggests that the model is relatively unresponsive to the true values. 

\edit{There are some caveats regarding the results presented in this section: (a) Due to the way that HSC PDR1 reserves PSF stars randomly for each exposure, 97\% of the stars in the PDR1 dataset were used to generate PSF models in more than one exposure before the coadding process \cite{2018PASJ...70S...5B}, so we are potentially underestimating the systematic uncertainties from the PSF interpolation process. (b) The results in this paper may overestimate $B[M_{04}]$ and $B[M_{40}]$ compared to the real HSC cosmic shear catalog, as the anti-correlation between $B[M_{04}]$ and $B[M_{40}]$ and seeing suggested that \textsc{PSFEx} severely overestimated $B[M_{04}]$ and $B[M_{40}]$ in good-seeing parts of the survey, which were eliminated from the shear catalog \citep{2018PASJ...70S..25M}.} \refresponse{Later HSC releases \cite{2022PASJ...74..247A} showed that the updated HSC coaddition method using the fifth-order Lanczos kernel did considerably better at modeling the PSF in good-seeing regions than the third-order Lanczos kernel in the first data release. Therefore, the modeling errors in the good-seeing fields are reduced for the later HSC three-year shear catalog \cite{2022PASJ...74..421L}. Given this resolution, we will not further investigate this particular issue. }


\section{Image simulation}
\label{sec:simulation}

In this section, we introduce the image simulations used in this study. The main purpose of the image simulation is to understand the shear response to the PSF higher moments modeling error, of which the methods and results are presented in this section. 

We will briefly cover the parts that are similar to the image simulation process in Section~3.3 of ZM21 
and focus on the details that are different from the previous paper. The general simulation workflow is introduced in Section~\ref{sec:simulation:workflow}, the galaxy profiles in Section~\ref{sec:simulation:galaxy}. In Section~\ref{sec:simulation:mom_method}, we introduce our method of manipulating PSF higher moments by changing the coefficients of the shapelet decomposition, and the PSF profiles used in this work in Section~\ref{sec:simulation:psf}. We show the results of the shear response to the PSF higher moment errors with image simulations in Section~\ref{sec:simulation:results}.

\subsection{Simulation Workflows}
\label{sec:simulation:workflow}

\begin{figure}
    \includegraphics[width = 1.05\columnwidth]{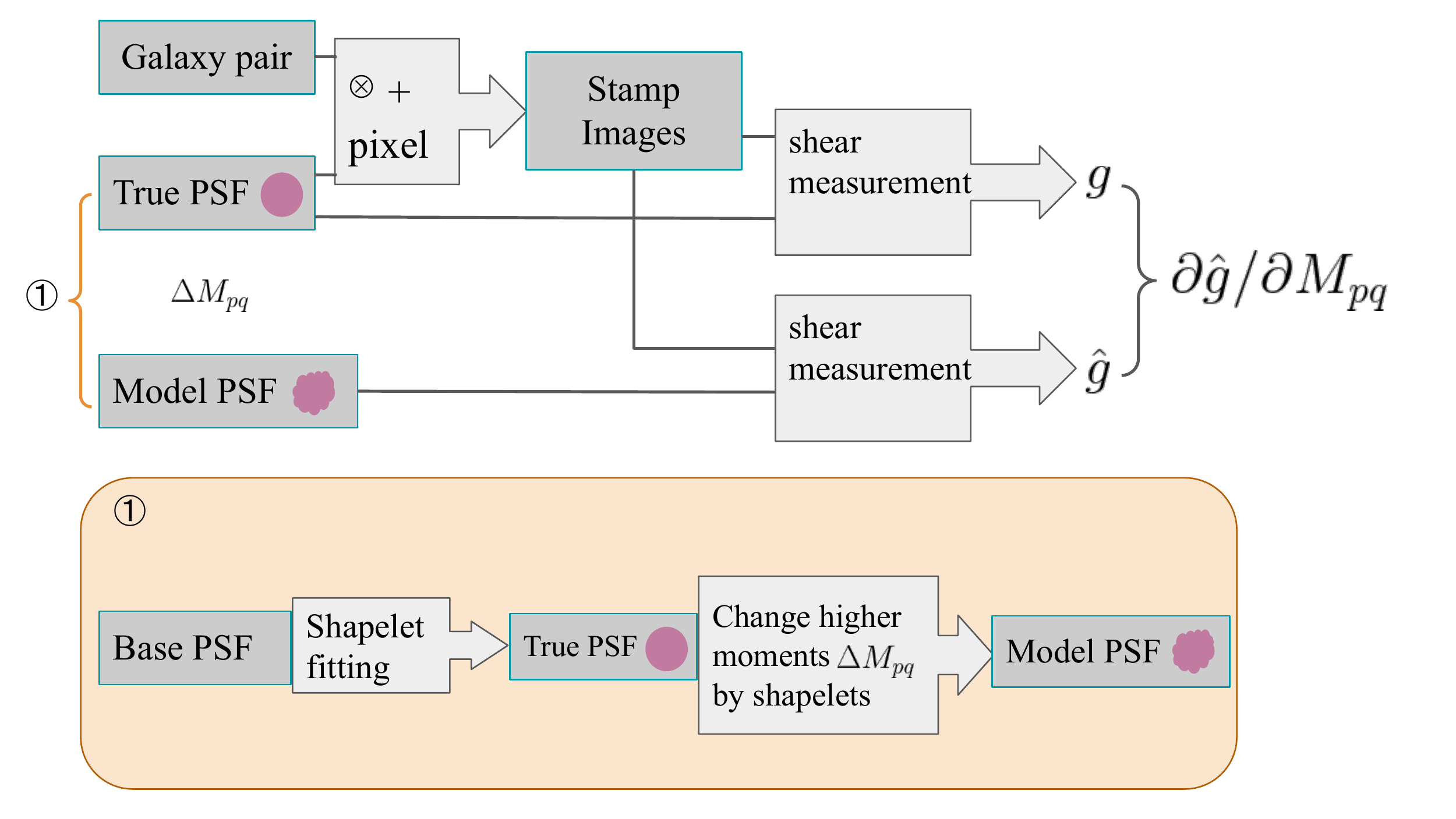}
    \caption{The workflow of the image simulation for one parametric galaxy and PSF model with one of the higher moment biased compared to the true PSF. The top part shows this workflow, while the bottom orange box shows the process that generates the true and model PSF. 
    }
    \label{fig:process_workflow}
\end{figure}

Fig.~\ref{fig:process_workflow} introduces the general image simulation workflow. The top part of the figure shows the steps of the image simulation process for one parametric galaxy and PSF. We started with a galaxy profile and its 90-\deg rotated pair \citep{Massey:2006ha}, an approach we used to reduce simulation volume by nullifying shape noise, for which the parameters will be introduced in Section~\ref{sec:simulation:galaxy}. The two galaxy profiles were convolved with the true effective PSF, introduced in detail in Section~\ref{sec:simulation:psf}; it includes the convolution with a pixel response function ($0.2$ arcsec). The convolved profiles were then sampled at the centers of pixels, generating the postage stamp images.  \edit{ The image set for the rotated galaxy pair was} fed into the shear measurement algorithm, which is the re-Gaussianization \citep{Hirata:2003cv} method implemented in the HSM module \citep{Mandelbaum:2005wv} in \textsc{GalSim} \citep{Rowe:2014cza}. We do not use Metacalibration \citep{Sheldon:2017szh,
Huff:2017qxu} as ZM21 showed that systematic biases in shear due to PSF modeling errors do not strongly depend on shear estimation methods.  We used the average of the measured shears for the galaxy and its 90-\deg rotated pair as the shear estimate for a given PSF. Finally, the difference between the two shear estimates $\Delta\hat{g}$, measured by the true PSF and the model PSF, provides the shear bias associated with the PSF higher moment bias $B[M_{pq}]$. 

The additive shear response to the higher moment error $M_{pq}$ was estimated at $g=0$ by
\begin{equation}
\label{eq:shear_response}
\frac{\partial c_{pq}}{\partial M_{pq}} = \frac{\Delta \hat{g}}{B[M_{pq}]}.
\end{equation}
To estimate the multiplicative shear bias generated by the PSF higher moment errors, we introduced another shear $g' = g + 0.01$. Its estimated values $\hat{g}'$ for the true and model PSF, and their  difference $\Delta \hat{g}'$, were used to estimate the multiplicative biases as
\begin{equation}
\label{eq:m_pq_estimate}
\frac{\partial m_{pq}}{\partial M_{pq}} = \frac{\Delta \hat{g}' -\Delta \hat{g}}{0.01B[M_{pq}]}.
\end{equation}

\refresponse{There are some general settings that apply to all of our image simulations:} we used \textsc{GalSim} \citep{Rowe:2014cza} to render the simulated images, all of which are noise-free postage stamp images with a pixel scale of $0.2$ arcsec, similar to the pixel scale of the Rubin Observatory LSST Camera (LSSTCam).

\subsection{Galaxy Profile}
\label{sec:simulation:galaxy}

\begin{table*}
\begin{tabular}{ccccccc}
\hline
Index & Galaxy Type & Galaxy Parameters & $(g_1, g_2)$                   & PSF Parameters                      & $B[M_{pq}]$                          \\\hline
1     & Gaussian    & $\sigma_{\text{gal}} = 0.17$ arcsec &  $(0, 0)$  & $\sigma_{\text{PSF}} = 0.24$ arcsec & $ -0.01 \sim 0.01$   \\
2 & Gaussian & $\sigma_{\text{gal}} = 0.17$ arcsec & $(0 \sim 0.01, 0 \sim 0.01)$  & $\sigma_{\text{PSF}} = 0.24$ arcsec & $0.005$ \\
3 & Gaussian    & $\sigma_{\text{gal}} = 0.1 \sim 0.9$ arcsec & (0.0, 0.0)                                                                                                                & $\sigma_{\text{PSF}} = 0.3$ arcsec & 0.005\\
4 & Sérsic, n=3 & $R_{\text{gal}} = 0.1 \sim 0.9$ arcsec      & (0.0, 0.0)                                                                                                                & $\sigma_{\text{PSF}} = 0.3$ arcsec & 0.005\\
5 & Gaussian    & $\sigma_{\text{gal}} = 0.1 \sim 0.9$ arcsec & $(0 \sim 0.01, 0 \sim 0.01)$                                                                                                                & $\sigma_{\text{PSF}} = 0.3$ arcsec & 0.005\\
6 & Sérsic, n=3 & $R_{\text{gal}} = 0.1 \sim 0.9$ arcsec      & $(0 \sim 0.01, 0 \sim 0.01)$                                                                                                                & $\sigma_{\text{PSF}} = 0.3$ arcsec & 0.005\\
7 & Bulge+Disc & $R_{h,b}, R_{h,d}, B/T, e_b, e_d$ in Table~\ref{tab:bpd_parameter}      & $(0 \sim 0.01, 0 \sim 0.01)$  & FWHM$ = 0.6$ arcsec & 0.005\\\hline
\end{tabular}
\caption{The specification of galaxies, PSFs, and higher-moments error applied to the PSFs for the single galaxy image simulations in this paper. The $M_{pq}$ in the last column stands for all viable moments from $3^{\text{rd}}$ to $6^{\text{th}}$ order.  
All base PSFs in the single galaxy simulations are Gaussian PSFs, except for the last row with Kolmogorov PSFs. Note that the PSF $\sigma$ values in the table describe the pixel-convolved true and model PSFs, not the base PSFs.
}
\label{tab:single_parameters}
\end{table*}

Two types of galaxy profiles were used in this study. The simpler galaxies were simulated as elliptical Gaussian light profiles. Gaussian galaxies were used in preliminary tests to develop basic intuition about the shear biases induced by errors in the PSF higher moments. The more complex galaxy model was a bulge+disc galaxy, consisting of a bulge and a disc component. The bulge+disc 
model was used for more sophisticated tests that attempt to represent a more realistic galaxy population as in the cosmoDC2 catalog \citep{2019ApJS..245...26K}.

The Gaussian profiles were parameterized by their size $\sigma$ and ellipticity $(e_1, e_2)$. We used them for initial tests to understand the relationship between shear bias and PSF higher moment bias (linear or non-linear?), the type of induced shear bias (multiplicative or additive?), and to determine which PSF higher moments actually contribute to weak lensing shear biases. 
The galaxy and PSF parameters for these preliminary single galaxy simulations are shown in Table~\ref{tab:single_parameters}, with results shown in Section~\ref{sec:simulation:results}. All base PSFs used in these initial simulations were Gaussian profiles, except for the last row, which is a Kolmogorov PSF.

A more sophisticated galaxy profile we used is the bulge+disc galaxy, a classic model used by many studies \citep[e.g.,][]{2006MNRAS.371....2A,2011ApJS..196...11S}. The bulges and disks in this work have common centroids. The bulge component was a de Vaucouleurs profile \citep{1948AnAp...11..247D}, a S\'ersic profile \citep{1963BAAA....6...41S} with $n = 4$, which means the surface brightness is proportional to $\exp(-R^{1/4})$, where $R$ is the distance from the centroid in units of its scale radius.  
The disk component was a\edit{n} exponential profile, i.e., the surface brightness is proportional to $\exp(-R)$\edit{, or the $n=1$ S\'ersic profile}. Both components have independent size and shape parameters. The luminosity profile of the components of the bulge+disc galaxy was governed by two parameters: total luminosity and the bulge fraction ($B/T$). The bulge+disc simulations allowed us to estimate the shear response to error in the PSF higher moments as a function of galaxy properties, which is an important input to the catalog-level simulations later in Section~\ref{sec:analyses:mock}.

\begin{figure}
    \centering
    \includegraphics[width=0.98\columnwidth]{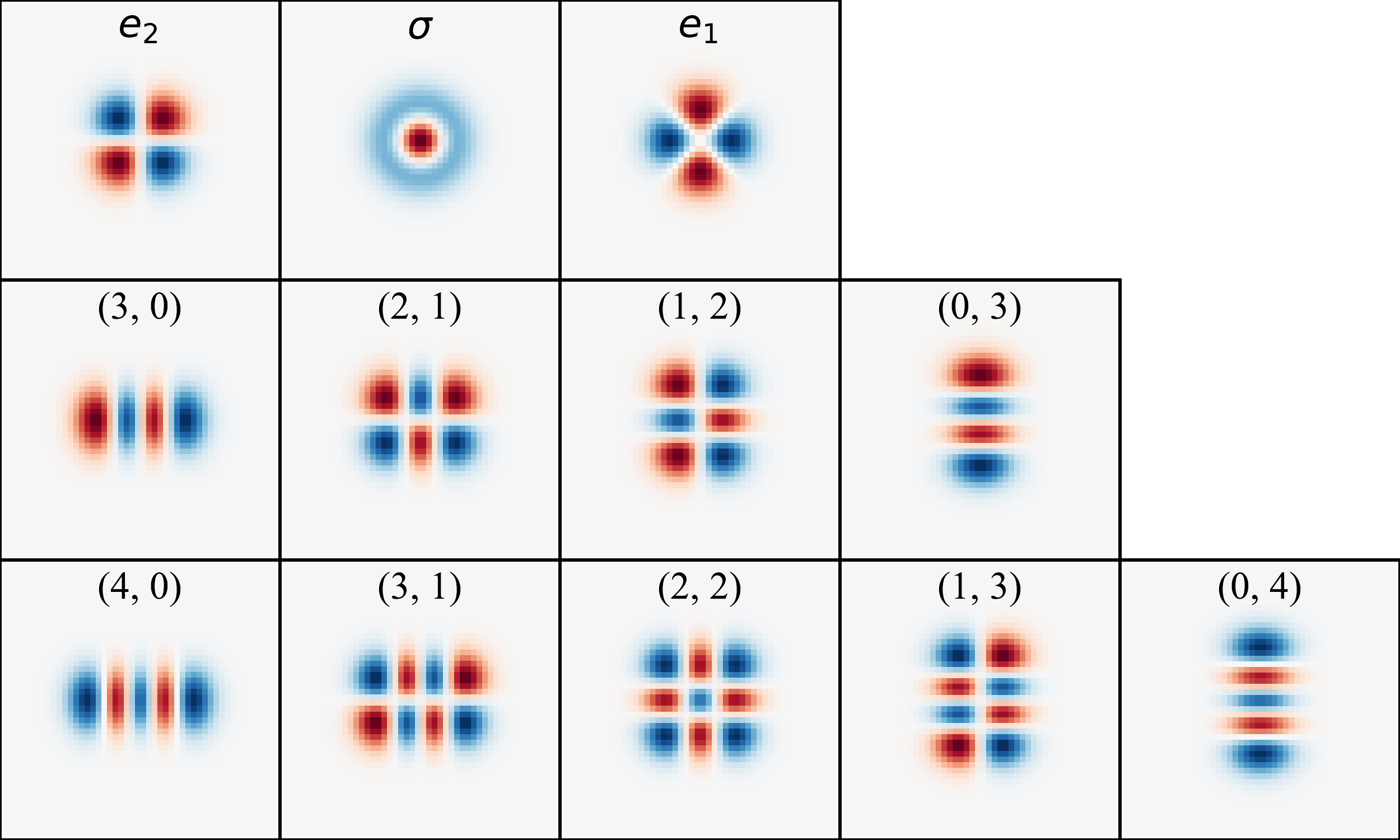}
    \caption{The moment \edit{responses} for a Gaussian PSF. We only show the second to fourth moments here, with index $(p,q)$ in Eq.~\eqref{eq:moment_define} labelled in each box. We use $e_1, e_2$, and $\sigma$ to represent the second moments. The color scale for each base covers $[-A,A]$, where $A$ is the maximum of the absolute value of the basis function. 
    }
    \label{fig:moment_basis}
\end{figure}

\subsection{Moment-Shapelet Relation}
\label{sec:simulation:mom_method}

Before introducing the PSF profile, we need a way to generate light profiles that differ in higher moments, introduced in Section~\ref{sec:background:mom_measure_mtd}, from the base PSF in ways that we can specify. Unfortunately, we do not know an analytical expression for a basis that has a one-to-one mapping with the higher moments. However, since the shapelet basis and the unknown \edit{moment response} can be used to describe the same linear space, we can reconstruct the unknown basis through linear combinations of the known shapelet basis, described in Section~\ref{sec:background:shapelet}. 

To do so, we defined the Jacobian matrix
\begin{equation}
\label{eq:jaco_def}
    T_{pq,jk} := \frac{\partial M_{pq}}{\partial b_{jk}},
\end{equation}
which is the generalized gradient of the moments $M_{pq}$ with respect to the shapelet coefficients $b_{jk}$ defined in Eq.~\eqref{eq:shapelet_expansion}.  We ranked the shapelet coefficients and PSF higher moments according to the orders in Fig.~\ref{fig:laguerre_basis} and Fig.~\ref{fig:moment_basis} 
We then directly estimated the change in moment $\Delta M_{pq}$ given the change in all shapelet coefficients $b_{jk}$,
\begin{equation}
\label{eq:linear_eq_1}
    \sum_{j,k} \frac{\partial M_{pq}}{\partial b_{jk}} \Delta b_{jk} = \Delta M_{pq}.
\end{equation}
Since $b_{jk}$ converges to zero at large $j+k$ for Gaussian-like profiles including ground-based PSFs, 
we were able to truncate the shapelet expansion at some finite order, making $\Delta b_{jk}$ and $T_{pq,jk}$ finite-sized vectors and matrices.

To numerically measure $T_{pq,jk}$ of the PSF with higher moment $M_{pq}$, we first decomposed the PSF into a set of shapelet coefficients $b_{jk}$. 
Then we perturbed $b'_{jk} = b_{jk} + \Delta b_{jk}$, and measured the higher moment $M'_{pq}$ after the perturbation. The Jacobian element was then estimated by
\begin{equation}
\label{eq:jaco_cal}
    T_{pq,jk} = \frac{M'_{pq} - M_{pq}}{\Delta b_{jk}}.
\end{equation}
In Appendix~\ref{ap:shapelet_moment}, we show a visualization of the Jacobian matrix that describes how PSF moments can be modified through changes in the shapelets coefficients. 

In the next section, we introduce the PSF profiles in this paper, and describe how we use the Jacobian $T_{pq,jk}$ defined in this section to precisely change the PSF higher moments.

\subsection{PSF Profile}
\label{sec:simulation:psf}

In the image simulations, we created the true and model PSF based on a ``base PSF''. We considered two base PSFs: Gaussian and Kolmogorov. Note that the base PSFs do not include the pixel response function, but the model and true PSFs do include it. 
The process to create the true and model PSF is shown in the orange box in Fig.~\ref{fig:process_workflow}.

To change the PSF moments using the technique described above, we first rendered an image of the base PSF including convolution with the pixel response function, and expanded that image by the shapelet decomposition implemented in \textsc{GalSim} \citep{Rowe:2014cza}.  
We carried out the shapelet decomposition up to order $10$, which corresponds to determining $66$ shapelet basis coefficients. To test that the shapelets decomposition is effectively representing the higher moments of the PSF profile, we confirmed that the fractional kurtosis error measured using the adaptive moments of the shapelets-reconstructed PSF compared to the original image is $10^{-5}$ for Kolmogorov and $10^{-9}$ for Gaussian, which is an acceptable precision for this study. The kurtosis is a good quantity for comparing higher moments, since (a) it is a combination of three moments ($M_{04}$, $M_{22}$, and $M_{40}$); (b) many other higher moments are zeros, and are not suitable for comparing fractional differences.

After representing the true PSF as an order $10$ shapelet series, we calculated the Jacobian $\boldsymbol{T}$ that links the $66$ shapelet coefficients with the PSF higher moments. The Jacobian is defined by Eq.~\eqref{eq:jaco_def} and estimated by Eq.~\eqref{eq:jaco_cal}. In this study, we investigated the higher moments from $3^{\text{rd}}$ to $6^{\text{th}}$ order, corresponding to $22$ moments. Together with the three second moments, the Jacobian is a $25 \times 66$ matrix. As an example, the Jacobian for the first 15 moments and first 15 shapelet modes is shown in Fig.~\ref{fig:Tij}.

Before describing how to use $\boldsymbol{T}$ to construct images with precisely modified higher moments, we first define our notation. The true and model PSF are represented as vectors of shapelet expansion coefficients $\mathbf{b}$ and $\mathbf{b'}$. The corresponding moment vectors are $\mathbf{M}$ and $\mathbf{M'}$.  

Ideally, we only change one higher moment of the PSF at a time, by solving for $\mathbf{\Delta b}$ in  Eq.~\eqref{eq:linear_eq_1}. 
However, because of the non-linearity of the moment-shapelet relationship, the higher moments will not change exactly according to $B[\mathbf{M}]$ when we add $\mathbf{b}$ and $\mathbf{\Delta b}$. Therefore, we introduced multiple iterations until the target moment biases $B[\mathbf{M}]$ are achieved, specified in Algorithm~\ref{alg:change_moment}. We defined $\mathbf{\Delta M}$ as the difference between our target moment vector and the current moment vector, which is the quantity we want to minimize. 
We used the L$^2$ norm to quantify the magnitude of $\mathbf{\Delta M}$, i.e., $||\mathbf{\Delta M}||_2 = \sqrt{\mathbf{\Delta M}^T \cdot \mathbf{\Delta M}}$.
\begin{algorithm}
\SetAlgoLined
 Initialize: $\mathbf{b}$, $\boldsymbol{T}$\;
 Target moment bias: $B[\mathbf{M}]$\;
 Target final moment vector: $\mathbf{M'} \leftarrow \mathbf{M} + B[\mathbf{M}]$\;
 $\mathbf{\Delta M} \leftarrow B[\mathbf{M}]$\;
 \While{$||\mathbf{\Delta M}||_2 > t_0$}{
  Solve $\boldsymbol{T} \mathbf{\Delta b} = \mathbf{\Delta M}$ for $\mathbf{\Delta b}$ \;
  Generate new model PSF: $\mathbf{\tilde{b}} =\mathbf{b} + \mathbf{\Delta b} $, measure its moments vector $\mathbf{\widetilde{M}}$ \;
  Update the $\mathbf{\Delta M}$: $\mathbf{\Delta M} \leftarrow \mathbf{M'} - \mathbf{\widetilde{M}} $ \;
  Update Jacobian: $\boldsymbol{T} \leftarrow  \frac{\partial \mathbf{\widetilde{M}}}{\partial \mathbf{\tilde{b}}}$
 }
 \caption{Moment Change}
\label{alg:change_moment}
\end{algorithm}

We used this algorithm to ensure that the moments of the new PSF model approach the target moments $\mathbf{M} + B[\mathbf{M}]$, so the new PSF model has moment biases that differ from those of the true PSF by $B[\mathbf{M}]$. 
We set the default threshold $t_0$ for the error in moment change to be $10^{-6}$, and the algorithm usually took less than 5 iterations to converge for Gaussian and Kolmogorov PSFs. Note that we included the second moments in the moment bias vector $B[\mathbf{M}]$  
and set them to zero. In this way, we \edit{actively verified} that the model and true effective PSF have the same second moments.

Introducing one component of $B[\mathbf{M}]$ at a time enabled us to inspect the \edit{moment response} from second to sixth order by taking the difference between the images before and after one moment is slightly biased, in Fig.~\ref{fig:moment_basis}. 
This also enabled us to quantify the impact on weak lensing shear associated with errors in the PSF model for a specific moment.

\subsection{Shear Response to PSF Higher Moments}
\label{sec:simulation:results}

\begin{figure*}
    \centering
    \includegraphics[width=1.7\columnwidth]{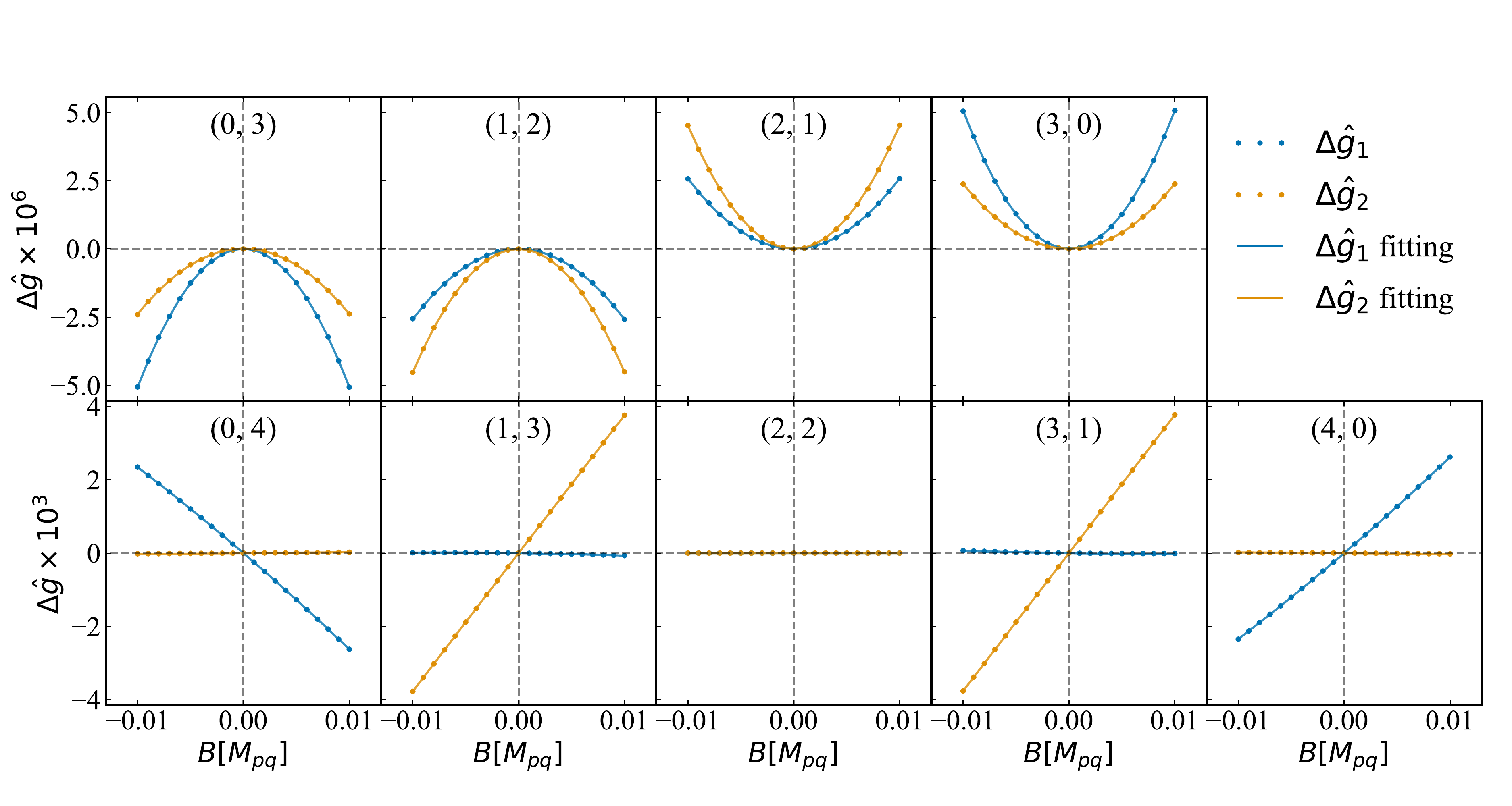}
    \caption{The additive shear bias generated by errors in the 3\textsuperscript{rd} and 4\textsuperscript{th} moments of the PSF. Both the galaxy and PSF have constant sizes. The shear biases for odd moments are well-fitted by a quadratic function, while those for even moments are linear. The quadratic fits are shown as lines,  
    while individual simulation results are shown by dots. The quadratic terms for the 4\textsuperscript{th} moments are $\approx 0$, so the fitting functions appear to be linear. As indicated in the y-axis labels, the order-of-magnitude difference in the additive shear biases between the 3\textsuperscript{rd} and 4\textsuperscript{th} moments is $10^3$.  } 
    \label{fig:shear_response}
\end{figure*}

\begin{figure*}
    \centering
    \includegraphics[width=1.7\columnwidth]{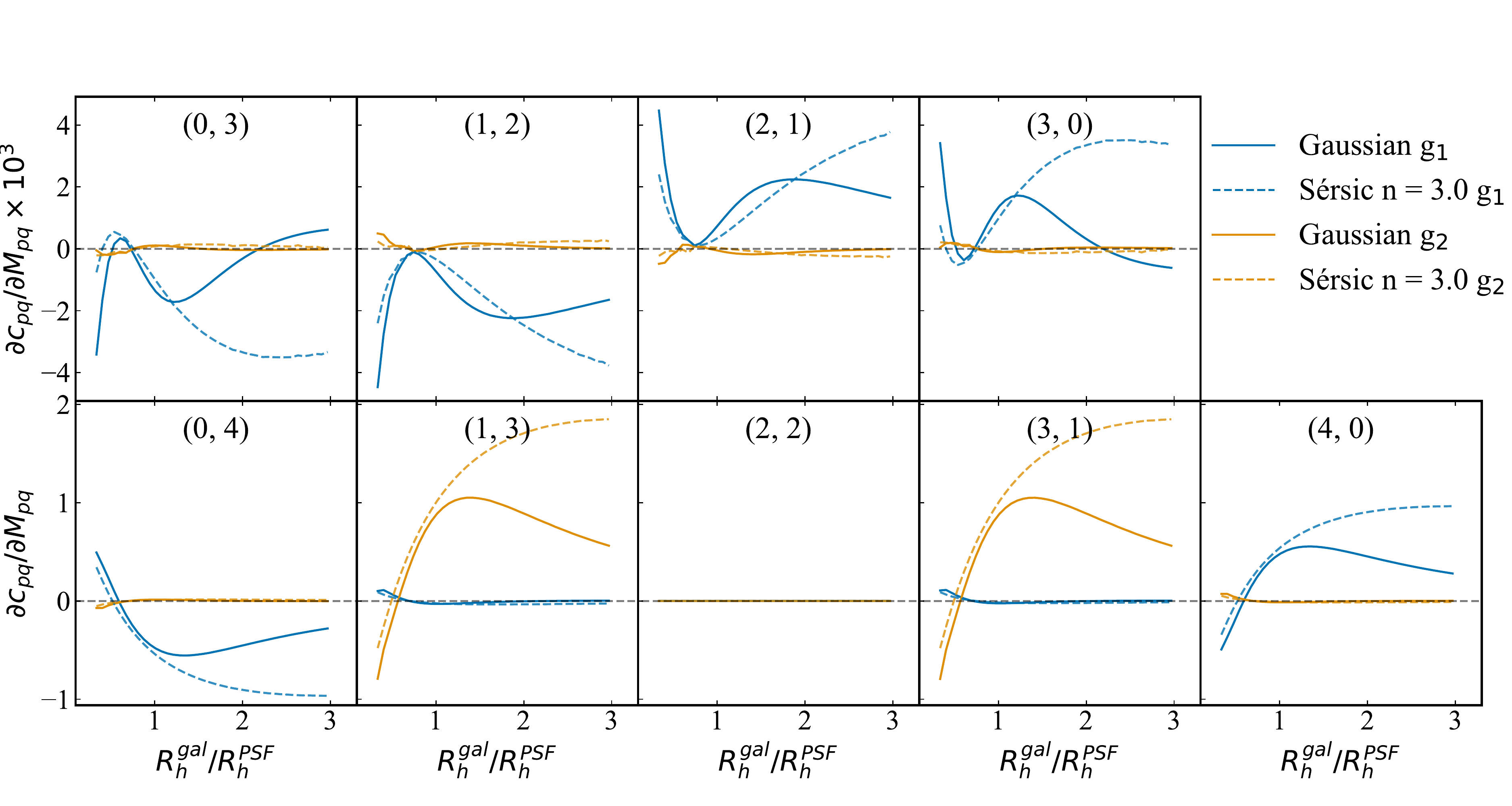}
    \includegraphics[width=1.75\columnwidth]{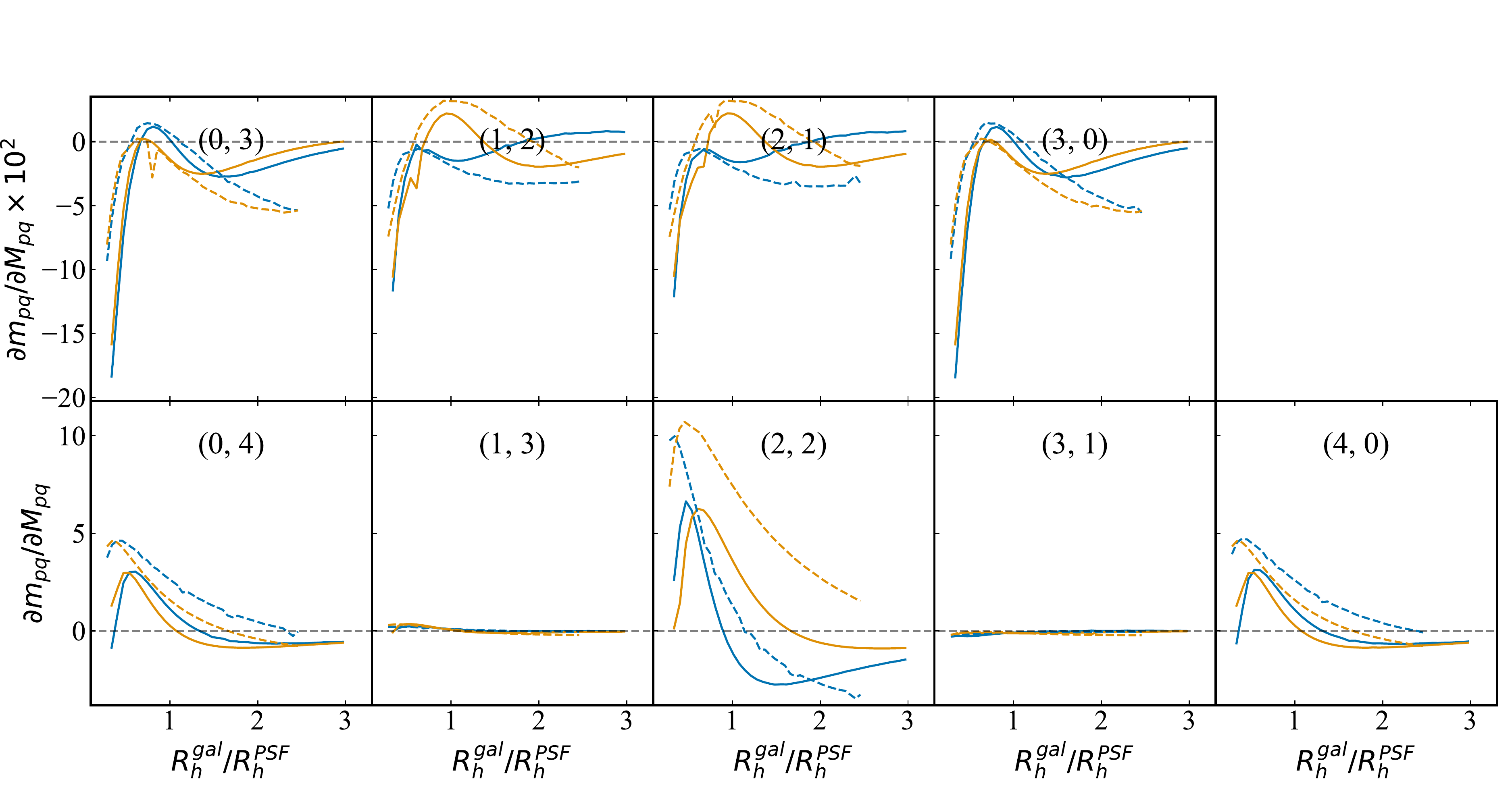}
    \caption{Additive (top) and multiplicative (bottom) bias responses to errors in the 3\textsuperscript{rd} and 4\textsuperscript{th}  
    PSF moments as a function of the ratio of the galaxy and PSF half light radii $R_{h}^{\text{gal}}/R_{h}^{\text{PSF}}$. We show results for both Gaussian galaxies and S\'ersic galaxies with $n = 3.0$, both with a Gaussian PSF. The size ratio is the primary factor determining the response, and the S\'ersic index of the galaxy is an important secondary parameter. As indicated in the y-axis labels, the order-of-magnitude differences in the additive (multiplicative) shear biases between the 3\textsuperscript{rd} and 4\textsuperscript{th} moments are $10^3$ ($10^2$).  
    }
    \label{fig:size_ratio_bias}
\end{figure*}

In this section, we show the results of the image simulation and shear measurement experiments described in Sections~\ref{sec:simulation:workflow} to \ref{sec:simulation:psf}, using Gaussian PSFs and 90-\deg rotated galaxy pairs. 
Using the single galaxy simulations, we can learn the following: (a) the form of the shear response to PSF higher moment errors -- are they linear, quadratic, or even more complicated; and (b) the pattern of shear biases associated with PSF higher moment errors, including magnitude of the biases and symmetry in the response to particular moments. Item (b) is particularly useful as it permits dimensionality reduction to focus on only the key PSF moments in later experiments. 

ZM21 found only multiplicative biases associated with the radial kurtosis error of the PSF model. In this study, we cannot assume that all biases will be multiplicative, since we introduced other moment errors. In Fig.~\ref{fig:shear_response}, we show the additive shear biases due to $B[M_{pq}]$ in the $3^{\text{rd}}$ and $4^{\text{th}}$ moments of the PSF model, with $(p,q)$ shown on top of each sub-plot.  The galaxy and PSF parameters are given in row~1 of Table~\ref{tab:single_parameters}. 
Fig.~\ref{fig:shear_response} shows that the $4^{\text{th}}$ moments induce shear biases that are linear in the moment residuals, while $3^{\text{rd}}$ moments induce shear biases that are non-linear in the moment residuals across the range of higher moment residuals seen in real data. We found that these curves can fit with a quadratic form. 
The shear response to the even moments is 2-3 orders of magnitude higher than to the odd moments, at a fixed $B[M_{pq}]$. 
We also note that the shear responses to conjugate higher moments, such as $M_{12}$ and $M_{21}$,  
have opposite signs. 
This is expected since the two moments are related through a 90-\deg rotation, causing an opposite effect on the shear. The symmetries in the shear responses to PSF higher moment errors are further discussed in Appendix~\ref{sec:app:symmetry}.  
To reduce the size of the figure, we omitted the 5\textsuperscript{th} and 6\textsuperscript{th} moments, but they exhibit the same trends as the 3\textsuperscript{rd} and 4\textsuperscript{th} moments in terms of parity symmetry and different order of magnitude between shear biases for odd and even moments. 

\begin{table}
\begin{center}
\begin{tabular}{llll}
\hline
Moment & $\frac{ m_{pq}}{ B[M_{pq}]}$ & $\frac{c_{pq}}{  B[M_{pq}]}$ & $\langle B[M_{pq}] \rangle \times 10^3$ \\\hline
(0,3)  & ($0.009, 0.001$)         & ($0.000, 0.000$)  &-0.21(0.24)     \\
(1,2)  & ($-0.005, 0.000$)        & ($0.000,0.000$)    &0.13(-0.04)     \\
(2,1)  & ($0.004,0.005$)          & ($0.000,0.000$)     &-0.07(-0.02)    \\
(3,0)  & ($0.002,0.000$)          & ($0.000,0.000$)     &0.34(-0.09)    \\
(0,4)  & ($2.223,1.550$)          & ($-0.255,0.002$)     &1.35(2.52)   \\
(1,3)  & ($-0.216,-0.166$)        & ($-0.005,0.376$)     &-0.01(-0.06)   \\
(2,2)  & ($1.940,5.367$)          & ($0.000,0.000$)      &-0.19(-0.16)   \\
(3,1)  & ($0.193,0.219$)          & ($-0.002,0.377$)     &-0.0(-0.04)   \\
(4,0)  & ($2.248,1.543$)          & ($0.255,-0.002$)     &1.02(3.67)   \\
(0,5)  & ($0.002,0.000$)          & ($0.000,0.000$)      &-0.96(0.86)  \\
(1,4)  & ($0.001,0.000$)          & ($0.000,0.000$)       &0.34(-0.13) \\
(2,3)  & ($0.003,0.005$)          & ($0.000,0.000$)       &-0.2(0.09) \\
(3,2)  & ($-0.001,0.005$)          & ($0.000,0.000$)      &0.33(-0.11)  \\
(4,1)  & ($0.001,0.002$)          & ($0.000,0.000$)       &-0.2(-0.13) \\
(5,0)  & ($0.000,0.000$)          & ($0.000,0.000$)       &1.5(-0.08) \\
(6,0)  & ($-0.360,-0.078$)          & ($0.110,-0.007$)     &3.42(11.77)   \\
(5,1)  & ($0.477,0.480$)          & ($-0.003,-0.206$)      &-0.05(-0.18)  \\
(4,2)  & ($0.072,-1.266$)          & ($0.105,0.028$)       &-0.16(0.49) \\
(3,3)  & ($0.029,0.012$)          & ($0.064,-0.413$)       &-0.02(-0.13) \\
(2,4)  & ($0.060,-1.95$)          & (-$0.105,-0.028$)      &-0.3(0.96)  \\
(1,5)  & ($-0.479,0.478$)          & ($-0.002,-0.206$)     &-0.02(-0.18)   \\
(0,6)  & ($-0.358,-0.071$)          & ($-0.110,0.008$)     &1.6(16.72)   \\
\hline      
\end{tabular}
\end{center}
\caption{Table of multiplicative and additive shear biases per unit of PSF higher moment residuals, $m_{pq}/ B[M_{pq}]$ and $c_{pq}/B[M_{pq}]$ , for the 3\textsuperscript{rd} to 6\textsuperscript{th} moments. Since the shear biases respond nonlinearly to the odd moment errors, values in this table are computed with the average PSF higher moment error of \textsc{PSFEx}, shown in Section~\ref{sec:data:hsc_measure}. \refresponse{We also list the mean of $B[M_{pq}]$ of the \textsc{PSFEx} (\textsc{PIFF}) in the last column for reference. }
}
\label{tab:mul_add_single}
\end{table}

Next, to measure both additive and multiplicative shear biases,  
we used the same galaxy and PSF sizes as in Fig.~\ref{fig:shear_response}, but we varied the lensing shear applied to the galaxies  (specified in row~2 of Table~\ref{tab:single_parameters}). In Table~\ref{tab:mul_add_single}, we show the multiplicative and additive shear biases per unit of PSF higher moments biases $m_{pq}/ B[M_{pq}]$ and $c_{pq}/ B[M_{pq}]$ for the 3\textsuperscript{rd} to 6\textsuperscript{th} moments, at the average PSF higher moment biases. Similar to Fig.~\ref{fig:shear_response}, the shear responses to the odd moments are at least two orders of magnitude smaller than the responses to the even moments. All even moments generate multiplicative shear biases, and they also strongly determine the additive biases. Notice that since the shear responds nonlinearly to the odd moments, the values for those moments in Table~\ref{tab:mul_add_single} depend on the PSF moment residuals. \edit{Based on the results from Section~\ref{sec:data:hsc_measure}, we can simply estimate the order of magnitude of $m$ and $c$ for a typical galaxy as being on the order of $10^{-3}$ to $10^{-2}$. A more precise estimate of the systematic biases for ensembles of galaxies will be provided in Section~\ref{sec:analyses:mock}.}

ZM21 showed that the galaxy-to-PSF size ratio is the most important factor that determines the shear response to the errors in modeling the PSF radial kurtosis.  Here we checked the sensitivity of the additive and multiplicative shear biases induced by individual PSF higher moment errors to that size ratio.  We explored this relationship by simulating Gaussian and S\'ersic galaxies with various sizes, specified in rows~3 to~6 in Table~\ref{tab:single_parameters}. In  Fig.~\ref{fig:size_ratio_bias},  we show the additive (multiplicative) shear biases in the upper (lower) panel, as a function of the galaxy-to-PSF size ratio measured by the half light radii $R_{h}^{\text{gal}}/R_{h}^{\text{PSF}}$. 
We can see that the size ratio plays an important role, but the S\'ersic index also affects the shear responses significantly, especially for large size ratios. This is consistent with the findings in ZM21.
\refresponse{In Fig.~\ref{fig:size_ratio_bias}, we note that the shear responses of Gaussian galaxies to the PSF third moments are non-monotonic, crossing the 0 reference line multiple times. The simulations in Fig.~\ref{fig:shear_response} corresponded to a galaxy-PSF size ratio of $0.7$, for which the third moment responses of $g_1$ and $g_2$ happen to have the same sign. As seen in Fig.~\ref{fig:size_ratio_bias}, the signs of the shear biases for the third moment residuals in Fig.~\ref{fig:shear_response} are not representative of many galaxy-to-PSF size ratios, and should not be over-interpreted. However, the small magnitude of the additive shear biases caused by third moment modeling errors in Fig.~\ref{fig:shear_response} are more generally applicable. }

In the next section, we will combine the findings in this section and in Section~\ref{sec:data} to estimate the systematic error in weak lensing observable and cosmology analyses associated with PSF higher moment errors.


\section{Weak Lensing and Cosmology Analyses}
\label{sec:analyses}

In this section, we discuss the propagation of errors in shear to the weak lensing 2PCF, and further into cosmology. We first provide a general derivation of our approach in Section~\ref{sec:analyses:general}, and then describe an important practical issue -- reducing the number of moments -- in Section~\ref{sec:analyses:reduction}. We introduce the mock galaxy catalog we use for estimating systematics, the cosmoDC2 catalog \citep{Korytov:2019xus}, in Section~\ref{sec:analyses:mock}. We further propagate the weak lensing shear systematics to cosmological parameter analysis using Fisher forecasts as described in Section~\ref{sec:analyses:fisher}. 

\subsection{General Error Propagation}
\label{sec:analyses:general}

Our discussion of how errors in the PSF higher moments affect the weak lensing 2PCF is based on two assumptions: (a) Each PSF higher moment may produce additive shear biases $c_{pq}$  and multiplicative biases $m_{pq}$ on the observed shear, $\hat{\gamma}  = (1+m_{pq}) \gamma +c_{pq}$.  (b) The total multiplicative and additive bias $m_{\text{total}}$ and $c_{\text{total}}$ produced by simultaneous errors in multiple higher moments of the PSF can be expressed as the sum of the individual multiplicative and additive biases $m_{pq}$,
\begin{align}
    \label{eq:add_property_1}
    m_{\text{total}} & \approx \sum_p \sum_q m_{pq}\\\label{eq:add_property_2}
    c_{\text{total}} & \approx \sum_p \sum_q c_{pq},
\end{align}
with uncertainties that are negligible for this work.  
The assumption (a) was illustrated in Section~\ref{sec:simulation:results}, 
and (b) was confirmed with an image simulation test, where 100 galaxies sampled from cosmoDC2 were assigned random PSF higher-moments residuals. That test showed that the absolute value of the differences between the two sides of Eqs.~\eqref{eq:add_property_1} and~\eqref{eq:add_property_2} for individual galaxies are $\le 10\%$. 
We have explicitly confirmed that for ensemble shear estimation, the error due to assumptions of linearity is further reduced to $\le 2$\%. 
For the multiplicative biases, since $m_{pq} \ll 1$, we can ignore the high-order correlations, and just focus on the first order expansion of the observed 2PCF of weak lensing shear. Additive biases can be written as the sum 
of their averages and fluctuations, $c_{pq}(\mathbf{x}) = c_{0, pq} + \tilde{c}_{pq}(\mathbf{x})$. Combining the additive and multiplicative terms, we get the full expression for the observed weak lensing 2PCF \edit{between bins $i$ and $j$},
\begin{align}
\label{eq:additive_deriv}
\langle \hat{\gamma}^i \hat{\gamma}^j \rangle = & (1 +  m_{\text{total}}(z_i) + m_{\text{total}}(z_j)) \langle  \gamma^i \gamma^j\rangle\\\nonumber 
    & +  \sum_{pq} \sum_{uv} \langle \tilde{ c}_{pq} \tilde{c}_{uv}\rangle +  c_{0,pq}  c_{0,uv},
\end{align}
\edit{where $m_{\text{total}}(z_i)$ is the multiplicative bias defined in Eq.~\eqref{eq:m0}. Throughout this work, we ignored the spatial variation of the multiplicative bias, which as shown by  \cite{2020OJAp....3E..14K} can enter the shear power spectrum at a lower level than the mean multiplicative bias. }

As shown in Eq.~\eqref{eq:additive_deriv}, the additive shear bias terms have two effects. First, the observed 2PCF is shifted by a constant $c_{0,pq}  c_{0,uv}$.  Second, it is also shifted by the scale-dependent auto-correlation function of the zero-mean additive bias field $\langle \tilde{c}(\mathbf{x}) \tilde{c}(\mathbf{x+\theta})\rangle$. We explore the impact of these changes in subsequent sections.

\subsection{Dimensionality Reduction for PSF Higher Moments}
\label{sec:analyses:reduction}

There are 22 correlated PSF moments from $3^{\text{rd}}$ to $6^{\text{th}}$ order, and the high dimensionality of this dataset can pose challenges in understanding the main issues determining the weak lensing systematic biases. Therefore, dimensionality reduction to only the PSF higher moments that induce substantial shear biases is an important first step. Since this task is based on a rough estimate of the importance of individual PSF higher moments, we used simple models for this: both the galaxy and PSF in the dimensionality reduction process are Gaussian profiles.

\begin{figure}
    \centering
    \includegraphics[width=1.0\columnwidth]{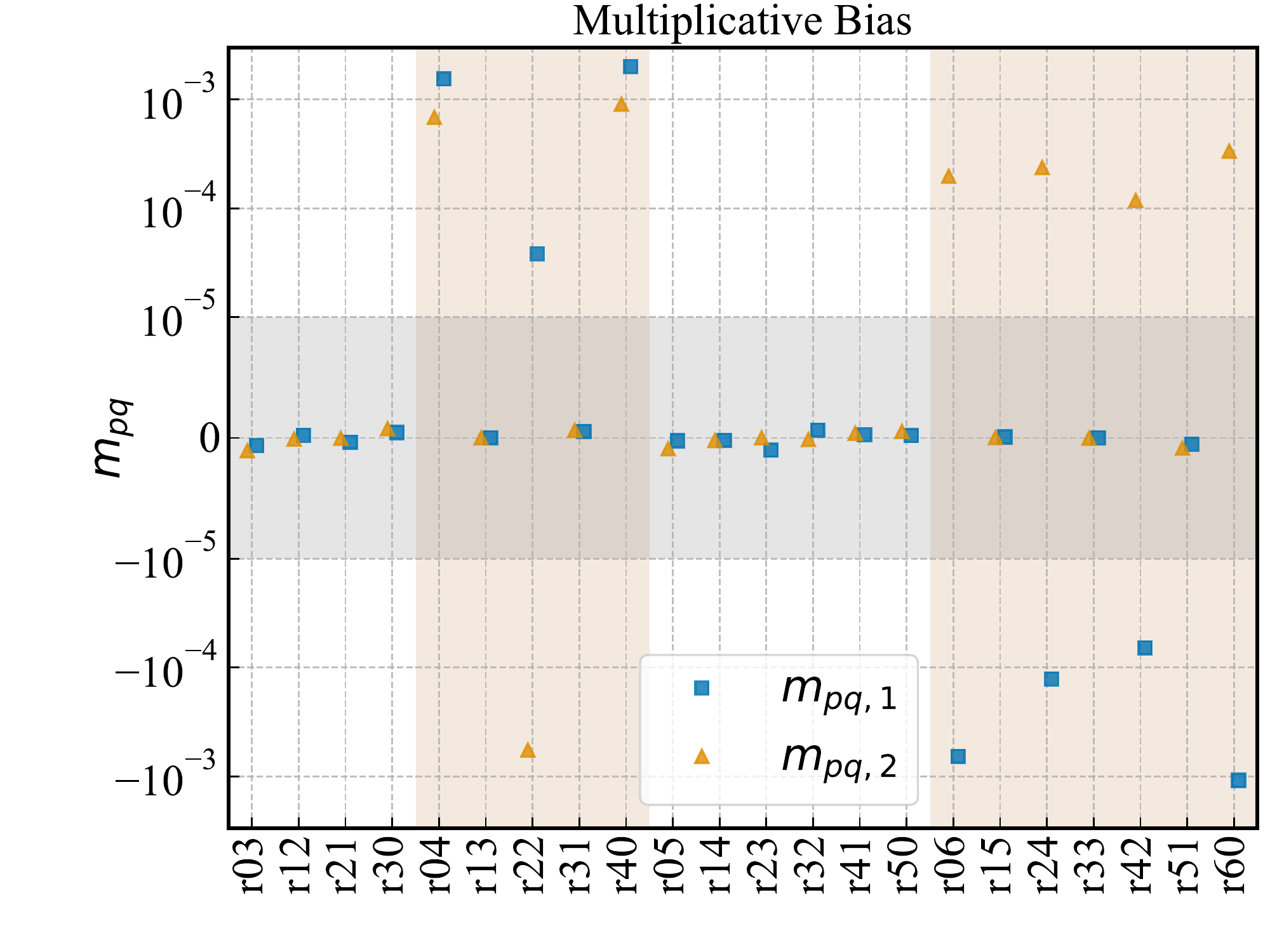}
    \caption{An estimate of the weak lensing shear multiplicative biases, aimed at understanding which PSF higher moments are most important in generating multiplicative biases.  This plot is based on ensemble shear biases for a simulated COSMOS galaxy sample, given the average error on individual  higher moments of the PSF model in HSC PDR1.  
    The orange areas are the even moments and the white areas are the odd moments. Both components of the multiplicative bias show the same set of 7 moments that contribute significantly. The y-axis is symmetrical log-scaled, with the grey area being the linear region. 
    }
    \label{fig:multi_prelim}
\end{figure}

Eq.~\eqref{eq:additive_deriv} shows that multiplicative bias affects the weak lensing 2PCF through its total $m_{\text{total}}$, which is a summation over all $m_{pq}$.  We used the methods described in Section~\ref{sec:simulation:workflow} to calculate $\partial m_{pq} / \partial M_{pq} (\sigma_\text{gal})$ as a function of the galaxy's second moment $\sigma_\text{gal}$. To roughly estimate $m_{pq}$, we used the $\sigma_\text{gal}$ of 44386 COSMOS galaxies with magnitude $< 25.2$, and galaxy resolution factor $R_2 > 0.3$ (later defined in Eq.~\ref{eq:resolution_factor}) as the input galaxy sizes. The second moments were computed after convolving with the Hubble PSF, but before convolving with our Gaussian PSF.
The Gaussian PSF size was fixed at a Full Width at Half Maximum (or FWHM) of $0.78~\text{arcsec}$. 
Assuming the shear bias is proportional to the PSF moment bias, the multiplicative bias should be proportional to the moment bias as well. Therefore, we estimated the multiplicative bias \edit{$\langle m_{pq} \rangle$ associated with $B[M_{pq}]$ as}
\begin{equation}
\label{eq:multiplicative_approx}
    \langle m_{pq} \rangle  = \left(\frac{1}{N}\sum_{i=1}^N \frac{\partial m_{pq}}{\partial M_{pq}} (\sigma_{\text{gal},i}) \right)\, \langle B[M_{pq}] \rangle,
\end{equation}
where the COSMOS galaxies are indexed by $i$, and $\langle B[M_{pq}] \rangle$ is the average moment bias of $M_{pq}$ in the HSC data, as described in Section~\ref{sec:data:hsc_measure}. The method to estimate $\partial m_{pq}/ \partial M_{pq}$ was described in Section~\ref{sec:simulation:workflow}.
We ranked the magnitude of the values of $\langle m_{pq} \rangle$ to estimate the importance of individual PSF moments. The importance is expected to be different for $g_1$ and $g_2$, given different spatial patterns are involved in different moments. 

The resulting multiplicative biases from this simplified simulation are shown in Fig.~\ref{fig:multi_prelim}. Both the $m_{\text{total},1} $ and $m_{\text{total},2} $ results indicate that PSF higher moments with both $p$ and $q$ even (seven in total) determine the multiplicative shear bias. The total multiplicative biases are $m_{\text{total},1} = 0.0017$ and $m_{\text{total},2}  = 0.0019$, dominated by the contributions of 7 PSF higher moments.

\begin{figure}
    \centering
    \includegraphics[width=1.0\columnwidth]{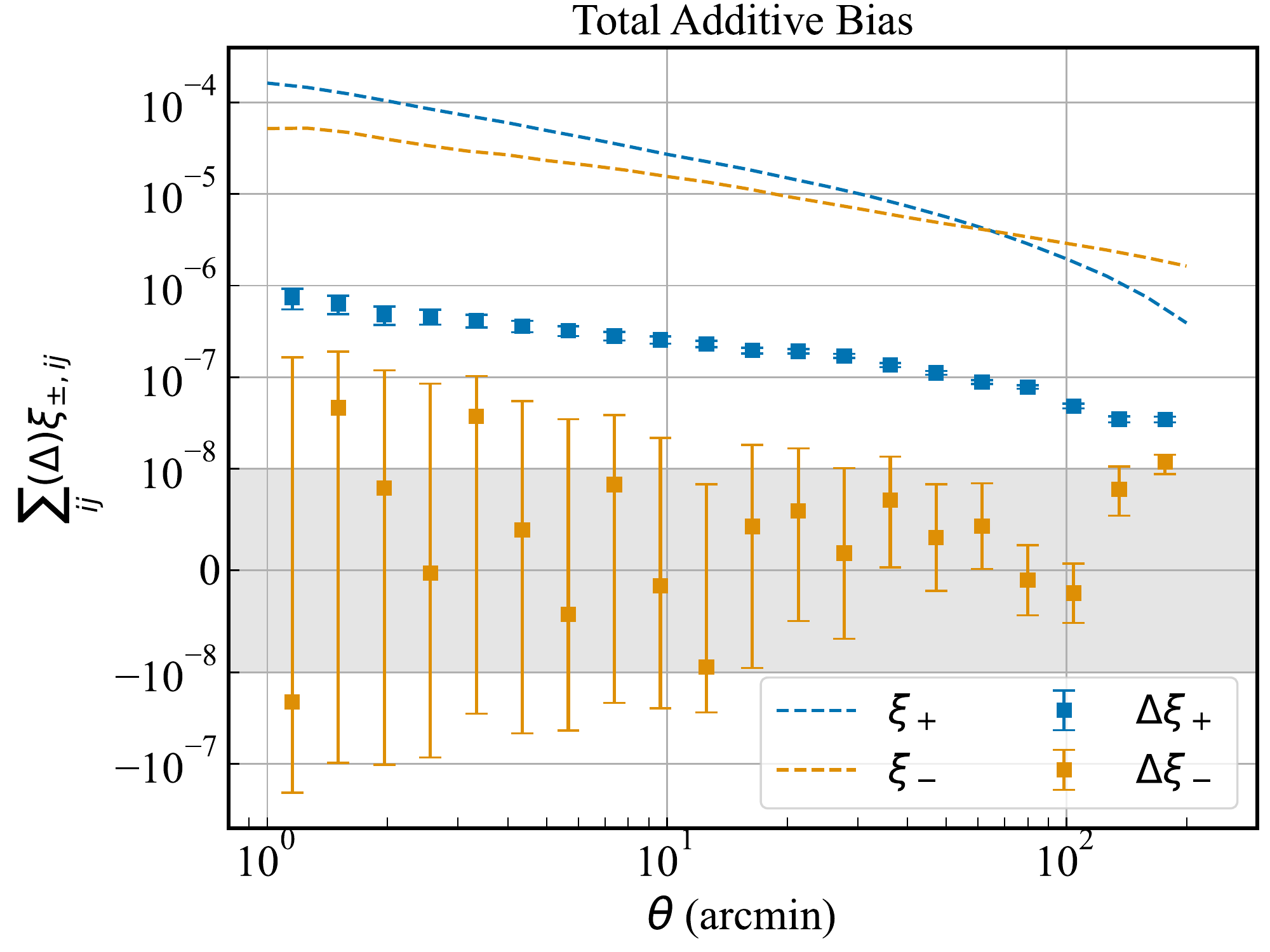}
    \caption{ The total additive bias on the weak lensing 2PCF $\xi_\pm$ for the simulated galaxies used for dimensionality reduction. The expected shear-shear correlation functions $\xi_\pm$ for our fiducial cosmological parameters (across all redshift bins combined) are shown as dashed lines. While $\Delta \xi_+$ is positive on all scales shown, $\Delta \xi_-$ is consistent with zero. 
    }
    \label{fig:two_groups}
\end{figure}

The additive biases are more complicated as shown in Eq.~\eqref{eq:additive_deriv}, since we must calculate the weak lensing 2PCF $\xi_{+/-}$ to understand the importance of the moments.  We designed the preliminary tests for the additive biases as follows: We used the PSF higher moments and their errors as a function of position in the HSC PDR1 from Section~\ref{sec:data}, and for the positions of bright stars in the PDR1 fields, we simulated a synthetic Gaussian galaxy with the average size and shape of the population from COSMOS catalog. We then measured the shear biases of the Gaussian galaxies with the PSF higher moments biases at these positions. We obtained the biases on the shear 2PCF directly from the shear bias at position $\mathbf{x}$, estimated by 
\begin{equation}
\label{eq:additive_approx}
    c_{pq} (\mathbf{x}) = \frac{\partial c_{pq}}{\partial M_{pq}} B[M_{pq}] (\mathbf{x}).
\end{equation}
As shown in  Fig.~\ref{fig:two_groups}, the additive bias on $\xi_+$ has a magnitude $\sim 10^{-7}$ on tens of arcmin scales, 
which corresponds to a $\sim 1$ per cent additive systematics contribution at small scales, and a few per cent at large scales, which is significant enough to potentially affect cosmological inference. The sharp decrease at $\theta \sim 100$ arcmin suggests that physical effects associated with the HSC field of view (FOV) are the cause of structural PSF systematic biases. However, $\Delta \xi_-$ is effectively zero.

\begin{figure}
    \centering
    \includegraphics[width=1.0\columnwidth]{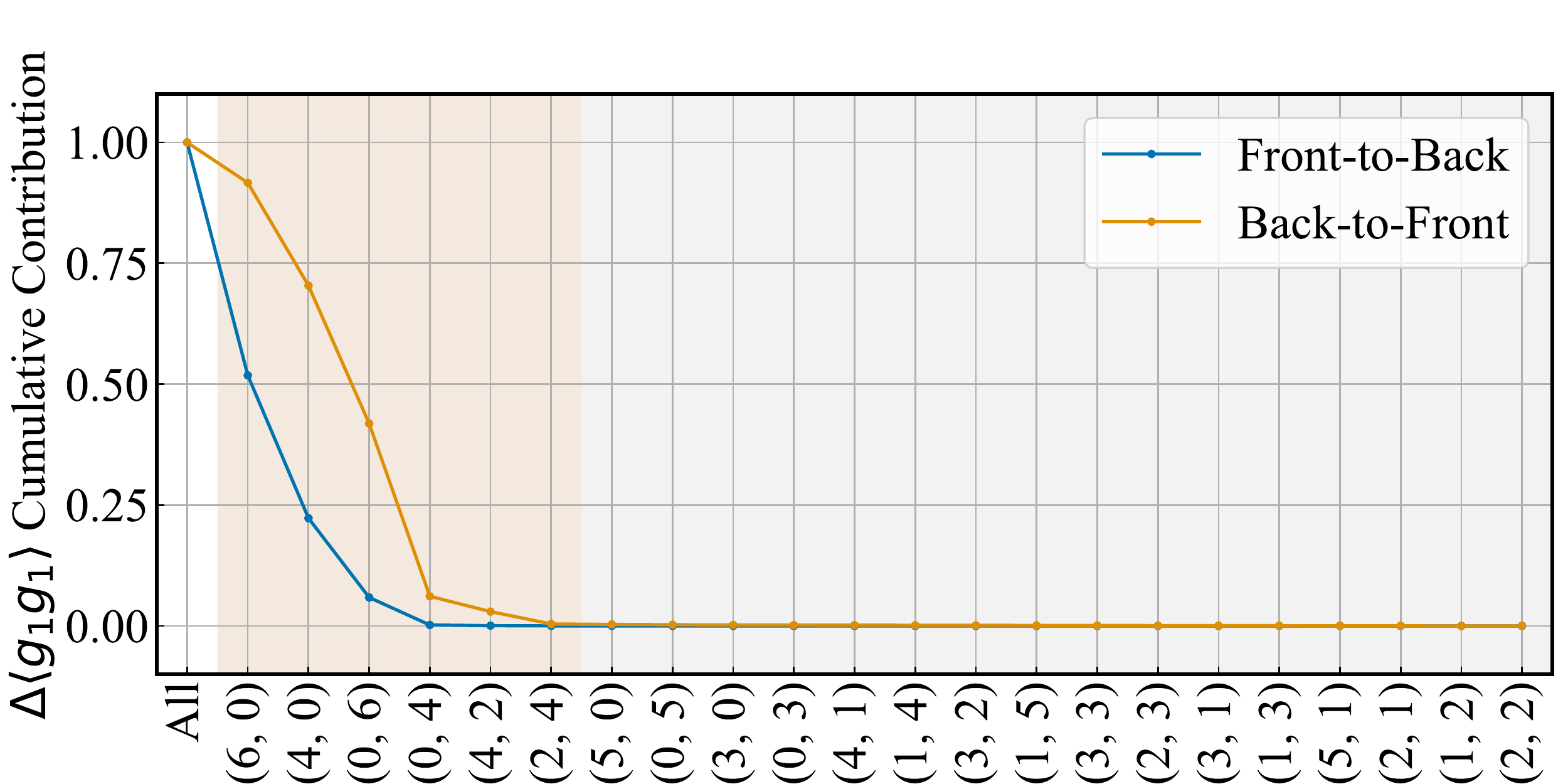}
    \includegraphics[width=1.0\columnwidth]{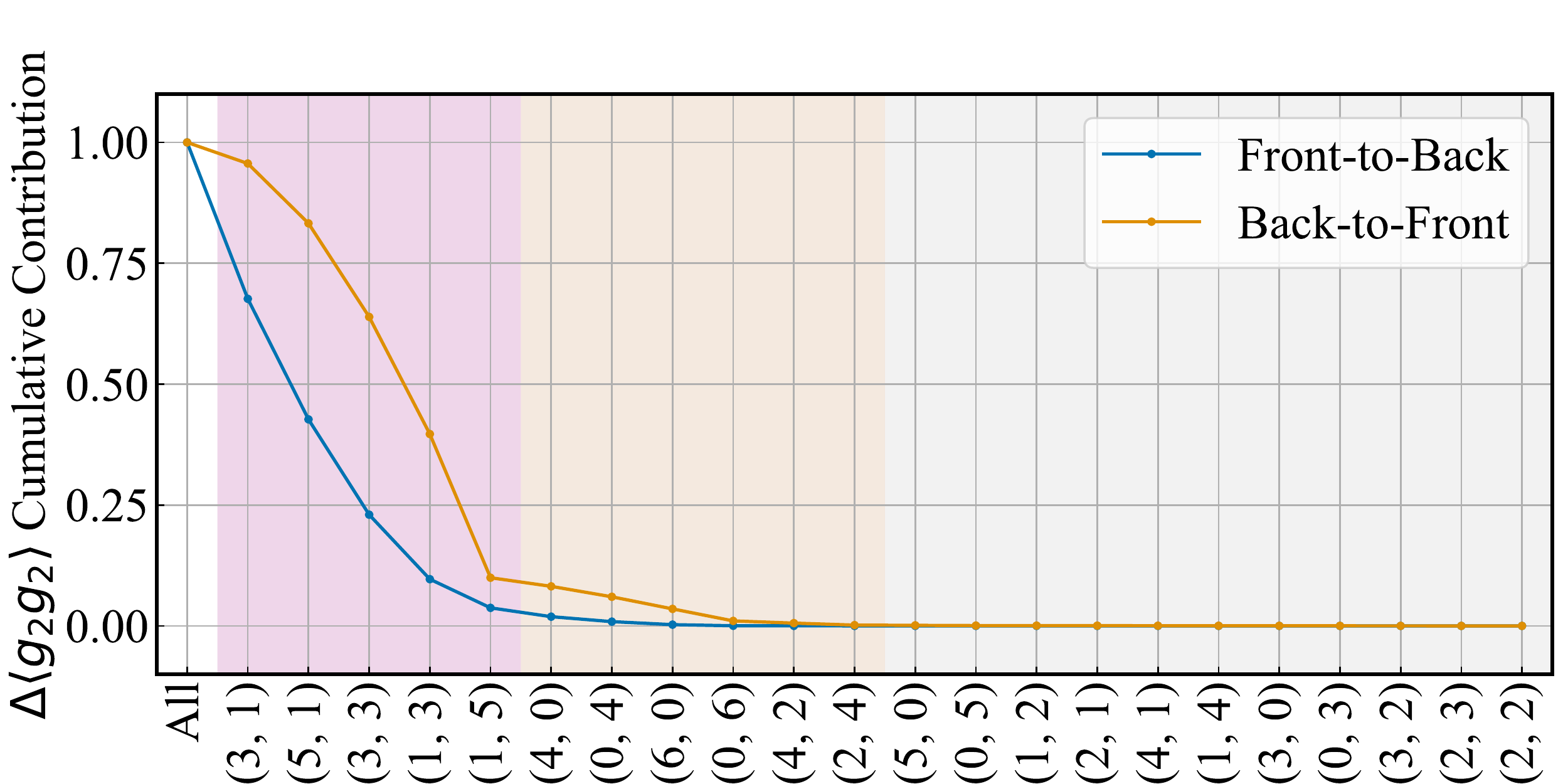}
    \includegraphics[width=1.0\columnwidth]{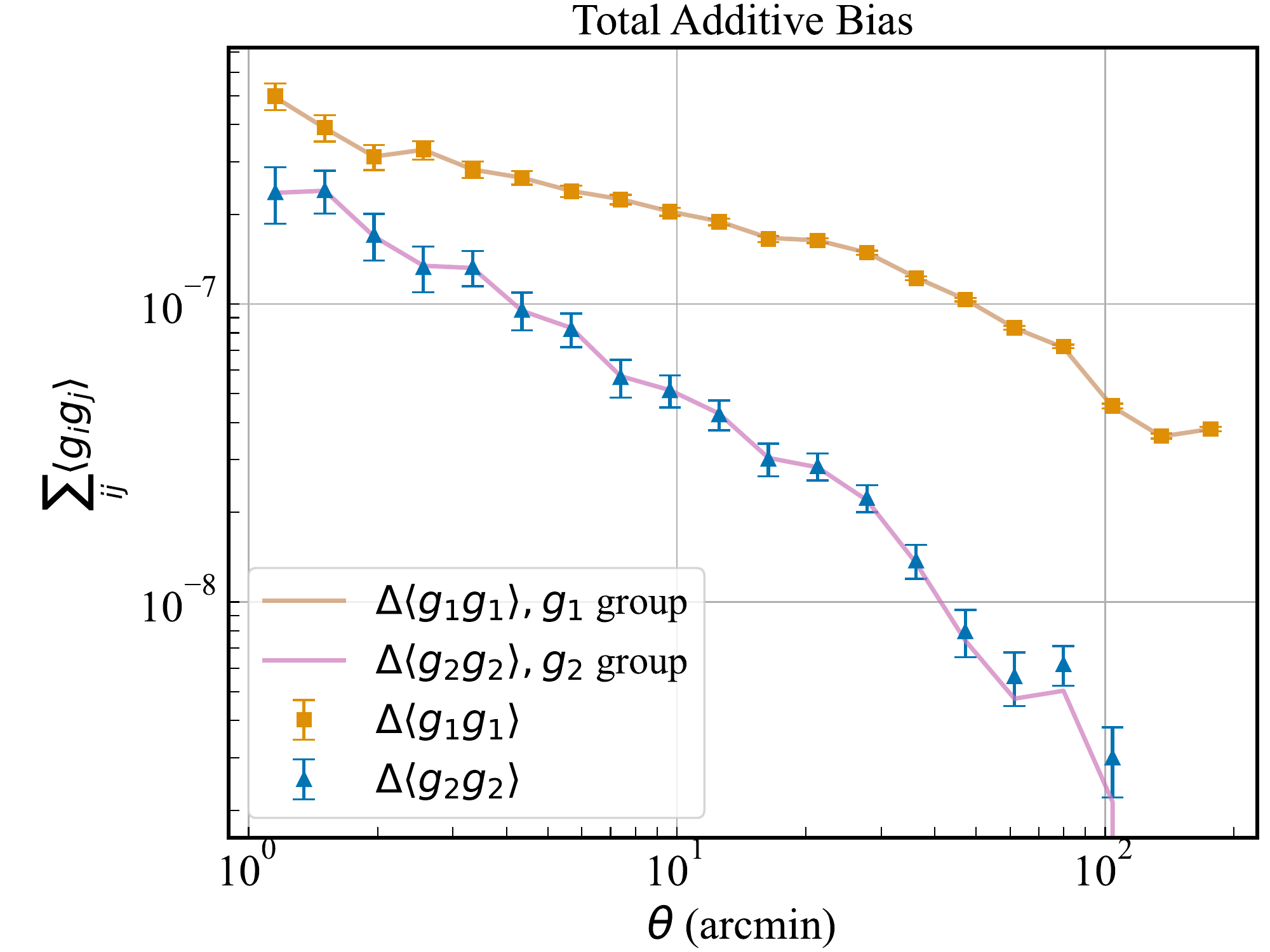}
    \caption{The estimate of the additive shear biases given the \textsc{PSFEx} modeling quality in the HSC PDR1.   The upper and middle panels show the rankings of the cumulative contribution to the $\Delta \langle g_1 g_1 \rangle$ and $\Delta \langle g_2 g_2 \rangle$ (respectively) from $2$ to $200$ arcmin, 
    from both the front-to-back and back-to-front methods described in Section~\ref{sec:analyses:reduction}. The light yellow region indicates the `$g_1$ group' moments that are most relevant to the $\Delta \langle g_1 g_1 \rangle$ term, and the pink region indicates the `$g_2$ group' moments that are most relevant to the $\Delta \langle g_2 g_2 \rangle$ term. The bottom panel shows the additive biases on $\langle g_1 g_1 \rangle$ and $\langle g_2 g_2 \rangle$ from all PSF higher moments, compared to just the `$g_1$ group' and the `$g_2$ group' -- confirming that these two groups dominate the additive shear biases.   
}
    \label{fig:add_prelim}
\end{figure}

Since each term in the additive biases on the 2PCF is associated with two different PSF moments (Eq.~\ref{eq:additive_deriv}), the ranking of importance for the PSF moments is more complex in this case. We designed two different ranking system: (a) the \textbf{front-to-back} approach and (b) the \textbf{back-to-front} approach. In the \textbf{front-to-back} approach, we calculated the contribution of each $\langle c_{pq} c_{uv} \rangle$ term to the total additive bias $\Delta \langle g g \rangle$, by integrating over $\theta$ from $1$ to $200$ arcmin. 
We ranked the contribution of a given moment $M_{pq}$ by the total reduction in additive bias if we removed all terms that involve $M_{pq}$. After removing the highest-contributing PSF moment, we performed the same calculation and removed the next highest-contributing moment, until only one moment remains.

Similarly, for the \textbf{back-to-front} approach, we removed the least-contributing PSF moment first, after performing the same contribution calculation described above. We then removed the next least-contributing moment, until we were left with only one moment. These two approaches provided two rankings of the PSF moments that contribute from most to least to the weak lensing additive shear bias. We expect to obtain a reasonably consistent set of PSF moments from these two approaches. 
If the two results were to disagree, the conservative approach would be to use the inclusive set of moments considered important by either method. 

We ranked the moments separately for $\langle g_1 g_1 \rangle$ and $\langle g_2 g_2 \rangle$. 
In Fig.~\ref{fig:add_prelim}, we show the results of executing the dimensionality reduction process for additive shear bias outlined in Section~\ref{sec:analyses:reduction}. In the upper and middle panel, we show the ranking of the PSF moments' contribution to $\langle g_1 g_1 \rangle$ and $\langle g_2 g_2 \rangle$. We show both the ``front-to-back method'' and ``back-to-front method'', described in Section~\ref{sec:analyses:reduction}. The relative rankings given by the two methods are slightly different, 
but the methods agreed about which moments we should discard. The moments that contribute the most strongly are four of the five $4^{\text{th}}$ moments: (4,0), (3,1), (1,3), (0,4), and all seven 
$6^{\text{th}}$ moments.  We further separated those 11 moments into two groups depending on which shear component they affect ($g_1$ or $g_2$). 
The moments in the $g_1$ ($g_2$) group are those with even (odd) values for both $p$ and $q$. In the bottom panel, we show the additive biases on $\langle g_1 g_1 \rangle$ and $\langle g_2 g_2 \rangle$ contributed by all PSF higher moments, compared to just the contributions of the `$g_1$ group' and the `$g_2$ group'. The plot shows that the `$g_1$ group' and `$g_2$ group' moments  dominate the total additive shear biases, and therefore we can focus on just these higher moments.

After the dimensionality reduction of PSF higher moments, we only propagate the errors on the reduced moment set to the lensing signal in the analysis in subsequent sections.  In other words, from this point on we only consider errors in 7 (11) PSF higher moments for the multiplicative (additive) biases.

\subsection{Mock Catalog Simulations}
\label{sec:analyses:mock}

To connect PSF higher moment errors with weak lensing systematics, we need a realistic galaxy catalog with galaxy properties and positional information. For this purpose, we used the cosmoDC2 catalog \citep{2019ApJS..245...26K}, as it is designed to match the galaxy population LSST is going to observe, with multiple validation tests against real datasets \citep{2021arXiv211003769K}, and has sufficient area ($\sim$440 \deg$^2$) for our purposes.  
We accessed the cosmoDC2 catalog using \textsc{GCRCatalogs}\footnote{\url{https://github.com/LSSTDESC/gcr-catalogs}} \citep{2018ApJS..234...36M}.

We estimated the multiplicative and additive shear biases for each individual galaxy in cosmoDC2 using two pieces of information: shear response to PSF higher moment errors, and a synthetic catalog of PSF higher moment errors, both described below. 


\subsubsection{Shear Response}
\label{sec:analyses:shear_response}

The shear response to errors in PSF higher moments, $\partial \hat{g} / \partial M_{pq}$, depends on the galaxy and PSF properties. 
We used a bulge+disc decomposition model for the galaxy, and determined the shear response as described in Section~\ref{sec:simulation}. To reduce the computational expense, we carried out simulations for a grid of bulge+disc model parameters that cover the majority of the cosmoDC2 galaxies, discarding $\lesssim 1$~per cent (large galaxies that do not contribute significant shear bias) outside of the grid.   
The free parameters in the grid are the half-light radius of the bulge $R_{h,b}$, the half-light radius of the disc $R_{h,d}$, and the bulge fraction $B/T$, and the grid is linear in all three dimensions. 
We used the same bulge and disc shapes for all galaxies\footnote{Our tests showed that using the same ellipticity for all galaxies generates $<$1 per cent error on the prediction of the ensemble shear biases, while saving tremendous computational time.}. We set the size and shape of the Kolmogorov PSF to be constant. The pixel size is $0.2$ arcsec, like that of the Rubin Observatory LSST Camera. The range of bulge+disc parameters in the image simulation is in Table~\ref{tab:bpd_parameter}.

\begin{table}
\begin{tabular}{ccc}
\hline
Parameter          & Range\\\hline
Bulge $R_{h,b}$       & $0.1 \sim 1.0~\text{arcsec}, \text{interval} = 0.1~\text{arcsec}$\\
Disc $R_{h,d}$       & $0.2 \sim 2.0~\text{arcsec}, \text{interval} = 0.2~\text{arcsec}$\\
Bulge-to-total ratio $B/T$  & $0.0 \sim 1.0, \text{interval} = 0.2$\\
Bulge shape & $e_1 = \pm 0.05, e_2 = \pm 0.05$\\
Disc shape & $e_1 = \pm 0.16, e_2 = \pm 0.16$\\
PSF FWHM & $0.6~\text{arcsec}$\\\hline
\end{tabular}
\caption{The parameters used in the bulge+disc image simulation. The top three rows define the parameter grid used for the simulation, while the bottom three rows are fixed parameters. We use the average absolute values of ellipticity for the bulges and disks. The $\pm$ signs indicate that the ellipticities of the galaxies in the 90-\deg rotated pairs have opposite signs.  The PSF FWHM shown is the size for the effective true and model PSFs.
} 
\label{tab:bpd_parameter}
\end{table}

After estimating a multiplicative and additive shear response to PSF higher moment errors $B[M_{pq}]$ at each grid point, we then used multi-dimensional linear interpolation from \textsc{SciPy}\footnote{\url{https://www.scipy.org/}} to estimate the multiplicative and additive shear biases for galaxies in cosmoDC2 using this grid. The \textsc{SciPy} routine performs a piece-wise interpolation in the 3-D parameter space\footnote{Our tests compared predictions for the ensemble shear bias of a sample of 100 simulated galaxies as estimated with the linear interpolation and with direct image simulations. We found no significant numerical difference between the two methods. }.

\subsubsection{PSF Moment Biases}
\label{sec:analyses:moment_generation}

Given the position for each galaxy in cosmoDC2, we need to assign PSF higher moment biases that reflect the average PSF higher moment biases and their correlation functions in the \textsc{PSFEx} dataset. Since cosmoDC2 is larger in area than any of the six HSC fields, it is impossible to directly cover the cosmoDC2 area with HSC fields. Therefore, we generated a synthetic PSF moment residual field $B[M_{pq}](x)$ with the same statistical properties as the \textsc{PSFEx} dataset, specifically the  average moment residuals and auto- and cross-correlation functions. The averages of the residuals are important for determining the multiplicative shear biases, and the correlation functions are important for the additive biases (see Section~\ref{sec:analyses}). 

As is described in Section~\ref{sec:data:hsc_measure}, the biases of PSF moments $M_{pq}$ and $M_{uv}$ are described by the average of the moment biases: $\langle B[M_{pq}] \rangle$,  $\langle B[M_{uv}] \rangle$, and the correlation function of the fluctuation $\xi^{pq,uv}(\theta)$. For the PSF moments that are of interest, we fit the correlation functions in the \textsc{PSFEx} dataset to parametric models and Hankel transformed them to get the angular power spectrum using \textsc{SkyLens}\footnote{\url{https://github.com/sukhdeep2/Skylens\_public/tree/imaster\_paper/}} \citep{2021MNRAS.508.1632S}, by computing
\begin{equation}
\label{eq:hankel_trans}
    C_\ell^{pq, uv} = 2 \pi \int \mathrm{d} \theta \,  \theta \, \xi^{pq,uv}(\theta) \, J_0(\ell, \theta), 
\end{equation}
where $J_0(\ell, \theta)$ is the Bessel function of order 0. Assuming the residual field is a Gaussian field, we generated the n-d correlated Gaussian field using these $n (n+1) /2$ angular power spectra. We used the python package \textsc{Healpy}\footnote{\url{https://github.com/healpy/healpy}} \citep{Zonca2019}, a python wrapper of the  \textsc{HEALPix}  
software\footnote{\url{http://healpix.sourceforge.net}} \citep{2005ApJ...622..759G},  
to generate a synthetic spherical harmonic decomposition $a_{\ell m}$ with $\ell_{\text{max}} = 3072$ and $-\ell \leq m \leq \ell$. With the $a_{\ell m}$, we generated an n-d Gaussian Random Field (GRF) evaluated at the centers of  \textsc{HEALPix} pixels with $N_{\text{side}} = 2048$, which corresponds to a pixel size of $1.7$ arcmin. The details of the GRF generation process are described in Appendix~\ref{ap:grf}. We then added the average moment biases for the \textsc{PSFEx} dataset to the GRF fluctuations to generate the total PSF higher moment bias fields. The PSF moment biases of any cosmoDC2 galaxy are the values for the \textsc{HEALPix} pixel that the galaxy sits in. The disadvantage of this method is that we cannot accurately evaluate $\langle\tilde{c}_{pq}(\mathbf{x})  \tilde{c}_{uv}(\mathbf{x+\theta})\rangle$ for angular bins below the \textsc{HEALPix} pixel size, i.e., $\theta \lesssim 1.7$ arcmin, though those scales make a negligible contribution to biases in cosmological parameters.

\subsubsection{Galaxy Selection and Weak Lensing Measurement}
\label{sec:analyses:selection}

The process outlined in the previous sections provided the galaxy responses $\partial \hat{g} / \partial M_{pq}$   
and the correlated PSF higher moment biases  $B[M_{pq}](\mathbf{x})$ for each galaxy in the cosmoDC2 catalog. However, not all of galaxies in this catalog will be used for lensing science in LSST. Similar to the practice in ZM21, we cut on how well-resolved a galaxy is based on its resolution factor $R_2$,  
which is calculated by
\begin{equation}
\label{eq:resolution_factor}
    R_2 = 1- \frac{T_P}{T_I},
\end{equation}
where $T_P$ and $T_I$ are the trace of the second moment matrix for the PSF and the galaxy, respectively. The galaxy is well resolved when $R_2 \rightarrow 1$, and poorly resolved when $R_2 \rightarrow 0$. Consistent with the approach used by the HSC survey  \citep{2018PASJ...70S..25M}, we only retained galaxies with $R_2 > 0.3$, eliminating $\sim$9 per cent of the sample\footnote{Since we did not simulate each cosmoDC2 galaxy, we estimated their resolution factors by interpolation from the galaxies on the grid.}.  
We excluded galaxies fainter than an i-band magnitude of $25.3$ for similar magnitude distribution as the LSST-`gold' samples \citep{2009arXiv0912.0201L}, and those outside the bounds of our grid of size values in Table~\ref{tab:bpd_parameter}. The lower limit of the size cut did not exclude any galaxies after the resolution factor cut, and the upper limit excluded $\sim 1 $ per cent of the galaxies.  After the cuts, the total number density of the catalog is $31.8$ arcmin$^{-1}$.

The bias on the 2PCF of the weak lensing shear $\Delta \xi_{+/-}$ was measured by
\begin{equation}
\label{eq:bias_2pcf}
\Delta \xi_{+/-}^{ij}(\theta) = \langle \hat{g}^i(x) \hat{g}^j(x+\theta) \rangle - \langle g^i(x) g^j(x+\theta) \rangle,
\end{equation}
where $i$ and $j$ are the tomographic bin index. In our measurement, we split the galaxies based on their true redshifts into three tomographic bins, centred at $0.5$, $1.06$, and $1.85$. 
The ensemble biases on the weak lensing 2PCFs $\Delta \xi_{+/-}^{ij}(\theta)$ were measured using \textsc{TreeCorr} \citep{2004MNRAS.352..338J}. 
In the next section, we use Fisher forecasts to understand the impact of these shear biases on cosmological parameter constraints. 

\subsubsection{Systematics on Shear 2PCF}
\label{sec:analyses:mock_results}

\begin{figure}
    \centering
    \includegraphics[width=1.0\columnwidth]{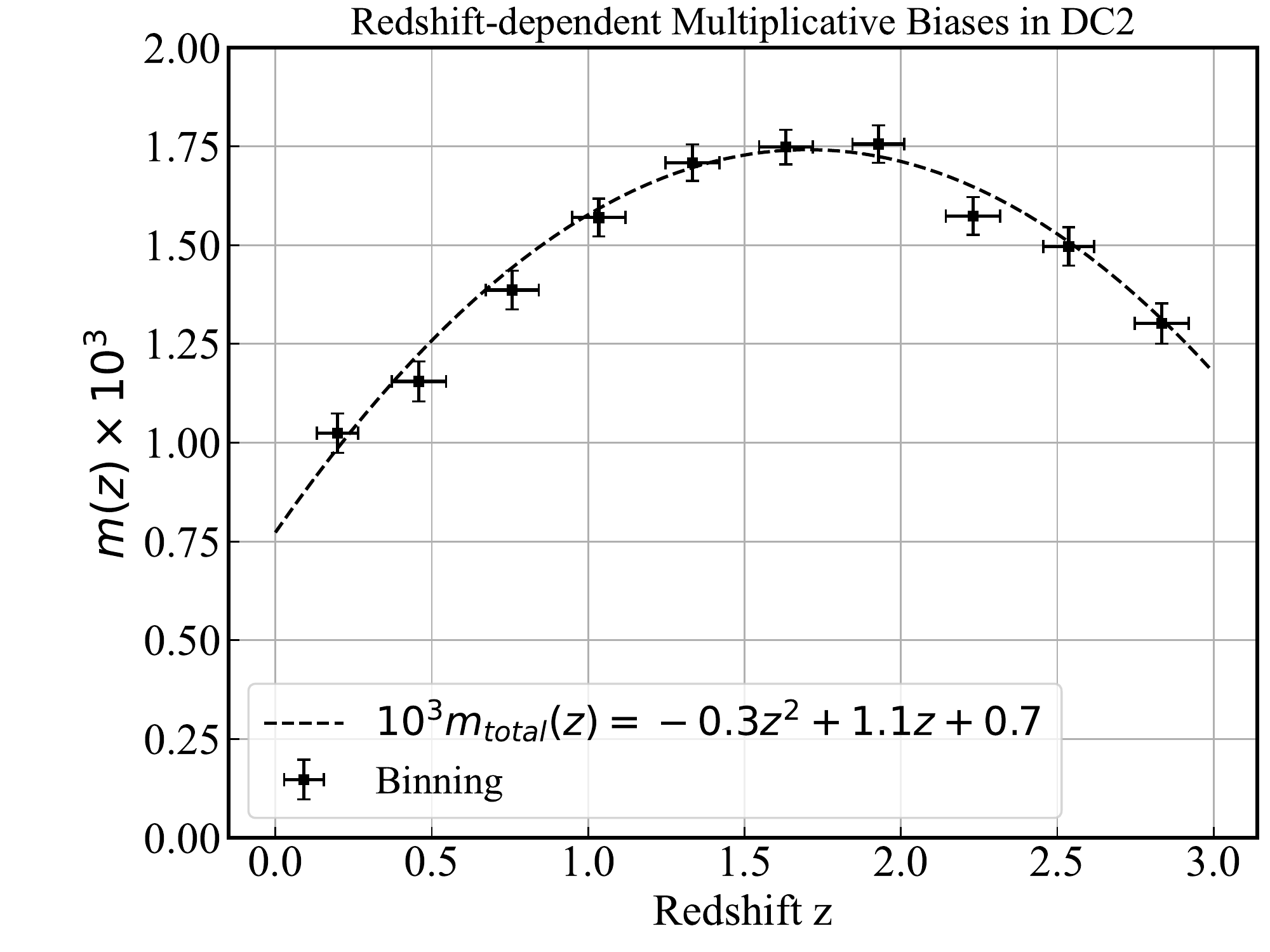}
    \caption{The redshift-dependent multiplicative shear biases for cosmoDC2  
    galaxies, due to PSF higher moment residuals comparable to those in HSC PDR1, in 10 redshift bins. We fit the data points to a quadratic function, shown as the dashed line. 
    }
    \label{fig:redshift_mul}
\end{figure}

In Fig.~\ref{fig:redshift_mul}, we show the total multiplicative biases of the cosmoDC2 galaxies in redshift bins after including all relevant PSF higher moment errors. We used a quadratic function to fit the 10 data points, and overplot the best-fitting curve as the dashed line. As suggested by \citet{2013MNRAS.429..661M}, a linear form for the redshift dependence of the multiplicative biases affects the estimate of the dark energy equation of state using weak lensing. 
The linear coefficient of our best-fitting $m(z)$ suggests that $m_0$ in Eq.~\eqref{eq:m0} is 0.0015, which is about half of the error budget in the LSST Y10 requirement \citep{2018arXiv180901669T}. Since the linear term of $m(z)$ can potentially cause significant cosmological parameter biases, and the impact of the quadratic term is unclear, we carried out a Fisher forecast for the impact of the redshift-dependent multiplicative biases, \edit{defined in Eq.~\eqref{eq:m0}}, on the inferred cosmological parameters, using the full quadratic $m(z)$. 

\begin{figure}
    \centering
    \includegraphics[width=1.0\columnwidth]{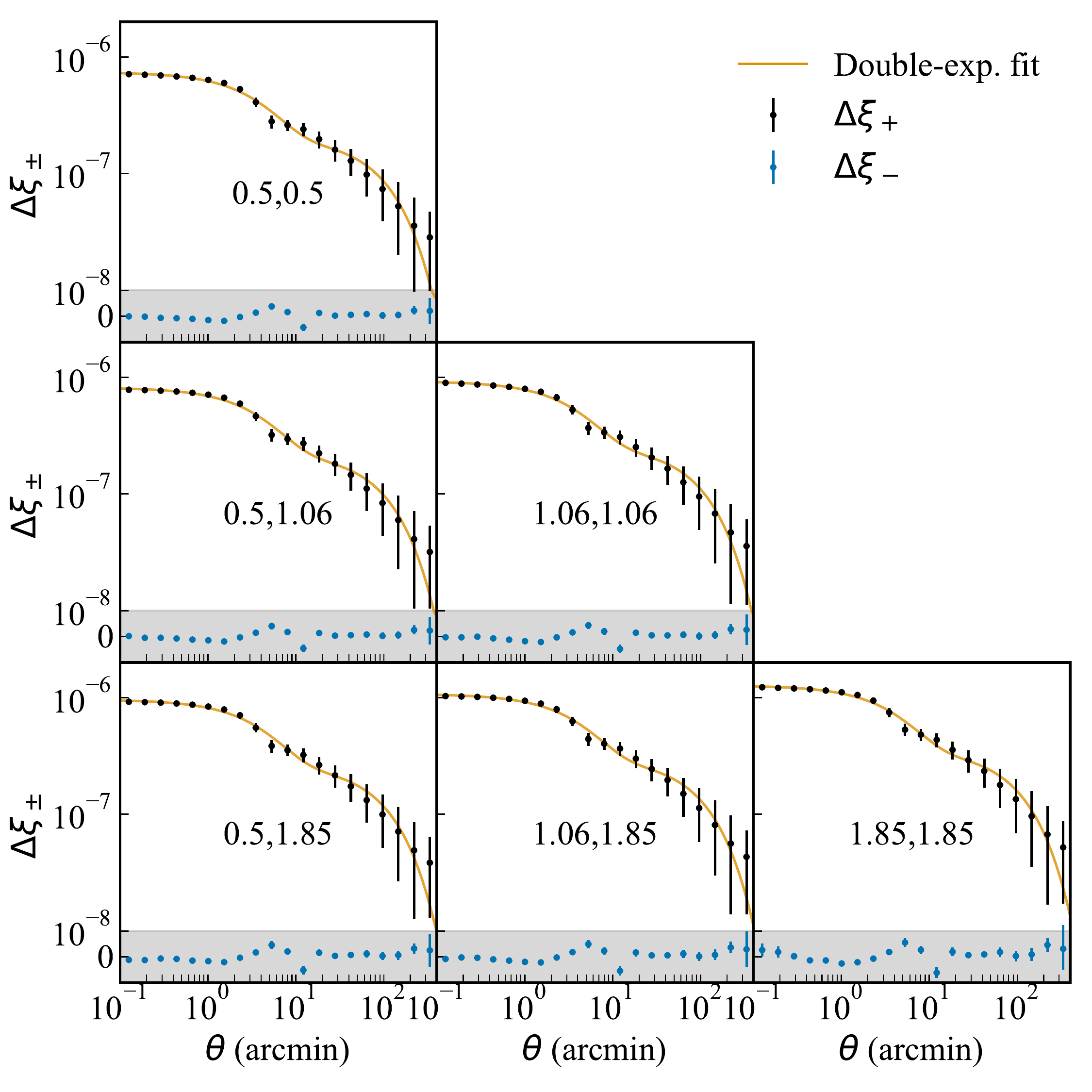}
    \caption{The additive biases on the weak lensing 2PCF $\xi_{\pm}$ for the cosmoDC2 galaxies when subjected to PSF higher moment residuals comparable to those in HSC PDR1. The galaxies are split into three tomographic bins based on their true redshifts, centred at $z=0.5$, $1.06$, and $1.85$. The tomographic bin combination is labeled by the central redshifts of the corresponding pair of bins at the center of each panel. The y-axis uses a symmetric-log scale, with linear scale $= 3.0 \times 10^{-7}$; the linear region is shaded grey. $\Delta \xi_-$ is still consistent with zero, as for the preliminary results. We fit $\Delta \xi_+$ with a double-exponential function, shown as orange lines. 
    }
    \label{fig:additive_tomographic}
\end{figure}

For the additive biases, we measured the difference in the weak lensing 2PCF, i.e., $\Delta \xi_+ = \sum_{pq} \sum_{uv} \langle \tilde{ c}_{pq}(\mathbf{x}) \tilde{c}_{uv}(\mathbf{x+\theta})\rangle +  c_{0,pq}  c_{0,uv}$, derived in Eq.~\eqref{eq:additive_deriv}.
In  Fig.~\ref{fig:additive_tomographic}, we show the additive biases $ \Delta \xi_{\pm}$, with galaxies split into three tomographic bins. Similar to the preliminary test, the additive biases on $\xi_+$ are positive, with magnitudes increasing at higher redshifts. $\Delta \xi_-$ is consistent with zero everywhere. We parameterized  $\Delta \xi_+$ as a double-exponential function, $\Delta \xi_+ = a_1 e^{-s_1 \theta} + a_2 e^{-s_2 \theta}$, as shown in orange.

In the next section, we propagate the estimated multiplicative and additive on the shear 2PCF, parameterized by the double-exponential function, to the cosmological parameter analysis using Fisher forecasts.

\subsection{Fisher Forecast}
\label{sec:analyses:fisher}

The goal of assessing the impact of PSF higher moment errors is to quantify their impact on a cosmological analysis using weak lensing shear, assuming that they are not explicitly accounted for in the analysis through modeling and marginalization. Since we only need an approximate estimate of the magnitude of induced cosmological parameter biases, we carried out a Fisher forecast on shear-shear data with 5 tomographic bins for the full LSST dataset (Y10).

In practice, we computed the Fisher information matrix elements $I_{ij}$ using the following equation:
\begin{equation}
    I_{ij} = \frac{\partial C_{\ell}}{\partial p_i}^T \text{Cov}^{-1} \frac{\partial C_{\ell}}{\partial p_j} + \delta_{ij} (\sigma_i \sigma_j)^{-1},
\end{equation}
where $i$ and $j$ are indices of the vector of parameters $\mathbf{p}$ (including both cosmological and nuisance parameters), $C_{\ell}$ is the angular power spectrum of the cosmic shear, and Cov$^{-1}$ is the inverse covariance matrix. The prior on each  parameter $p_i$ was added to its diagonal element in the Fisher information matrix as $1/\sigma_i^2$, where $\sigma_i$ is the standard deviation of the Gaussian prior.
We used the DESC Science Requirements Document (SRD) covariance matrix \citep{2018arXiv180901669T}.

The forward model in this forecast includes 7 cosmological parameters ($\Omega_m$ the matter density, $\Omega_b$ the baryonic matter density, $h$ the Hubble parameter, $n_s$ the spectral index, the power spectrum normalization parametrized as $\sigma_8$ and the dark energy equation of state parameters $w_0$ and $w_a$), 4 intrinsic alignment (IA) parameters of the non-linear alignment model \citep[NLA;][]{2017MNRAS.470.2100K}, i.e., the IA amplitude $A_0$, redshift-dependent power-law index $\eta_l$, redshift-dependent power-law index at redshift $z > 2$ $\eta_h$, and luminosity dependent parameter $\beta$.  
The Fisher forecast code and setup was adapted from and explained more thoroughly in Almoubayyed, et al., {\em in prep}. The fiducial values and priors of all parameters are shown in Table~\ref{tab:cosmology_parameters}. 

\begin{table}
\centering
\begin{tabular}{cccccc}
\hline
Parameter           & Value & Prior $\sigma$  & Parameter & Value & $\sigma$ \\\hline
$\Omega_m$          & 0.3156 & 0.2 & $A_0$ & 5.0 & 2.0    \\
$\sigma_8$       & 0.831 & 0.14  & $\eta_l$ & 0.0 & 2.0      \\
$\Omega_b$  & 0.049 & 0.006     & $\eta_h$ & 0.0 & 2.0   \\
$h$ & 0.6727 & 0.063  & $\beta$ & 0.0 & 2.0\\
$n_s$ & 0.9645 & 0.9645   &  &  &   \\
$w_a$ & 0.0 & 2.0  & & & \\
$w_0$ & -1.0  & 0.8 & & &\\\hline
\end{tabular}
\caption{The fiducial values of and priors on the cosmological and intrinsic alignment parameters we use as the baseline of the Fisher forecasting. 
}
\label{tab:cosmology_parameters}
\end{table}

Derivatives of the angular power spectrum with respect to these parameters were taken using \texttt{numdifftools} \citep{numdifftools} with an absolute step-size of 0.01, which was validated to be stable through a convergence test in  Almoubayyed, et al., {\em in prep}, 
and for the cosmological parameters, was also shown to be stable in \citet{2021arXiv210100298B}.  

The $C_{\ell}$ values were computed in 20 $\ell$ bins, consistent with the binning used in the DESC SRD, using the Core Cosmology Library \citep{CCL}. The additive shear 2PCF biases for the tomographic weak lensing signal for redshift bins $i$ and $j$ measured in cosmoDC2 were parameterized by
\begin{equation}
\label{eq:fitting_function}
\Delta \xi_+^{ij} (\theta) = a_1^{ij} e^{-s_1^{ij} \theta} + a_2^{ij} e^{-s_2^{ij} \theta},
\end{equation}
where the parameters $a_1^{ij}$, $a_2^{ij}$, $s_1^{ij}$, and $s_2^{ij}$ are linear functions of $z_i + z_j$, the sum of the mean redshifts of the tomographic bins being correlated. This fitting function was empirically selected based upon visual inspection, and all fractional fitting residuals are within $3\%$ of the true values. 
\edit{Using the fitting function in Eq.~\eqref{eq:fitting_function} enables us to calculate the 2PCF additive biases for any tomographic binning.}

The model for the additive biases associated with PSF higher moment errors has in total 8 parameters. The multiplicative biases were modeled for each tomographic bin, using a quadratic function to fit $m(z)$. Our model for the 2PCF with multiplicative biases is 
\begin{equation}
\label{eq:multiplicative_tomo}
\hat{\xi}_+^{ij} = (1+m^i(z_i) + m^j(z_j))\xi_+^{ij},
\end{equation}
where $\hat{\xi}_+^{ij}$ and $\xi_+^{ij}$ are the observed and true cosmic shear 2PCFs. Since the multiplicative shear biases for individual bins were determined from a quadratic fitting formula,  only 3 parameters are needed to model the multiplicative biases. The 2PCF additive biases for the 15 tomographic bin-pairs were calculated using the best-fitting parameters for the linear functions of $z_i + z_j$. 
Next, they were Hankel transformed to obtain biases in the angular power spectra, $\Delta C_\ell$. The forecasted biases on the cosmological and intrinsic alignment parameters $p_i$ were calculated using \citep{10.1111/j.1365-2966.2005.09782.x}
\begin{equation}
\label{eq:param_bias}
B[p_i] = \sum_j (I^{-1})_{ij}  \frac{\partial C_{\ell}}{\partial p_j}^T  \text{Cov}^{-1}  \Delta C_{\ell}.
\end{equation}
We compared the bias $B[p_i]$ on each parameter with its forecasted 1$\sigma$ uncertainties from the Fisher matrix formalism in order to determine the relative importance of the systematic biases on cosmological parameter constraints due to PSF higher moment errors, if not corrected or removed.

\begin{figure}
    \includegraphics[width=1.0\columnwidth]{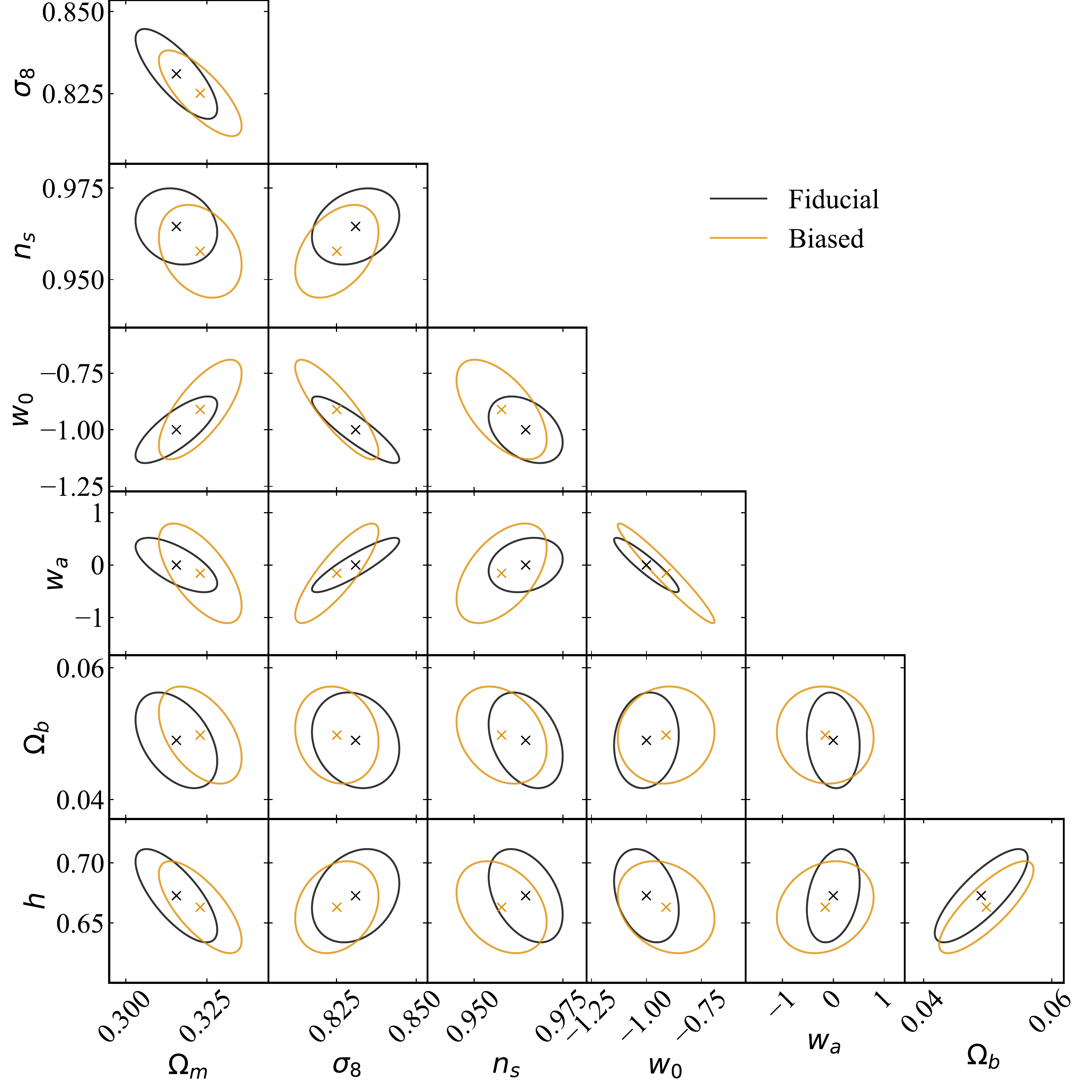}
    \caption{The $1\sigma$ constraint contours from the Fisher forecast for the fiducial (black) and shifted by additive shear biases (orange) cosmological parameters for LSST Y10. The centroids of the forecasts are shown by the ``x''. If not accounted for in the analysis, the additive shear biases caused by errors in the PSF higher moments at the level produced by \textsc{PSFEx} for HSC PDR1 are predicted to shift the inferred cosmological parameters by $\sim 1\sigma$, at the LSST Y10 level.  
    }
    \label{fig:xi_fitting}
\end{figure}

In Fig.~\ref{fig:xi_fitting}, we show the cosmological parameter shifts induced by failure to account for the additive shear biases caused by PSF higher moment residuals when interpreting cosmic shear measurements at the level of LSST Y10 \citep{2018arXiv180901669T}.  In this forecast, we marginalized over the intrinsic alignment parameters $A_0$, $\beta$, $\eta_l$, and $\eta_h$. The shifts in cosmological parameters $B[p_i]$ caused by errors in the PSF higher moments correspond to $\sim 60$ to $\sim 100$ per cent of their $1\sigma$ uncertainties. 

Next, we applied redshift-dependent multiplicative biases $m(z)$, shown in Fig.~\ref{fig:redshift_mul}, to the cosmic shear $\xi_\pm$ in the Fisher forecasts. For LSST Y10 \citep{2018arXiv180901669T}, we found that these multiplicative biases only shift the cosmological parameters by a few per cent of their $1\sigma$ uncertainties. 
As discussed in Section~\ref{sec:analyses:mock_results}, the linear coefficient of $m(z)$ suggests that we have $m_0 = 0.0015$ in Eq.~\eqref{eq:m0},  
which corresponds to around $50$ per cent of the systematic error budget for this parameter. This prediction overestimates the impact of the redshift-dependent multiplicative biases on the cosmological parameter estimates compared to our Fisher forecasts. The most likely reason for this finding is that our $m(z)$ is dominated by the quadratic term rather than the linear term, and therefore the redshift-dependent multiplicative shear bias is less degenerate with structure growth than the linear shear bias in Eq.~\eqref{eq:m0}.

We repeated the Fisher forecast analysis for LSST Y1, incorporating differences in its redshift distribution and covariance matrix. 
The LSST Y1 forecast yielded a larger $\sigma$ for all of the parameters $p_i$. For the additive biases, our analysis predicted that the average $|B[p_i]|/\sigma$ for LSST Y1 is 0.21, compared to 0.73 for LSST Y10, over the parameters that the cosmic shear constrains, i.e., $\Omega_m$, $w_0$, $w_a$, and $\sigma_8$. For the multiplicative biases, our analysis predicted that this average $|B[p_i]|/\sigma$ for LSST Y1 is 0.039, compared to 0.062 for LSST Y10. In general, the PSF higher moment errors affect the results for LSST Y1 less so than LSST Y10, but they still must be accounted for in the Y1 analysis, if the PSF modeling is not improved. 

In summary, our Fisher forecast analysis showed that the PSF higher moment errors of \textsc{PSFEx} as applied to HSC PDR1 (if not reduced in magnitude or marginalized over in the analysis) can cause up to a $1\sigma$ shift in the cosmological parameter estimates in an LSST Y10 cosmic shear analysis.  This result is dominated by additive biases; the multiplicative biases only shift the estimated cosmological parameters by $\sim 0.1\sigma$ according to the Fisher forecast.


\section{Conclusions and Future Work}
\label{sec:conclusions}

In this paper, we have presented the results of a comprehensive study of the weak lensing shear biases associated with errors in modeling the PSF higher moments (beyond second moments) for ground-based telescopes, following the previous path-finding paper that identified the potential for non-negligible weak lensing systematics due to this effect for LSST (ZM21). We have quantified the additive and multiplicative shear biases due to errors in the 3\textsuperscript{rd} to 6\textsuperscript{th} moments of the PSF, including 22 moments in total, including estimating the typical magnitude of these errors when using current PSF modeling algorithms, and propagating them to the impact on cosmological parameter estimation. 

To carry out this study, we developed an  iterative algorithm that uses a shapelet expansion to modify individual PSF  moments in our image simulations while preserving the other moments. Using this approach, we measured the multiplicative and additive shear responses, $\partial m_{pq} / \partial M_{pq}$ and $\partial c_{pq} / \partial M_{pq}$, to the individual PSF moment errors. We identified trends in these quantities with the galaxy-to-PSF size ratio and the S\'ersic index of the galaxy. The behavior of the shear responses can be summarized as follows: 

\begin{enumerate}
    \item Given the typical magnitude of modeling errors in PSF higher moments, the amplitude of the shear biases due to errors in the odd moments of the PSF is 2-3  magnitude smaller than those caused by the even moments, which means that they can be ignored. 
    \item For the even moments, the multiplicative and additive shear biases are linear functions of the moment biases $B[M_{pq}]$, and the responses primarily depend on the galaxy-to-PSF size ratio and S\'ersic index. 
    \item Other galaxy parameters, e.g., bulge fraction $B/T$ and galaxy shapes, play a more minor role in determining the shear biases due to PSF higher moment errors. 
    \end{enumerate}

As an example of the current state of the art, we have measured the modeling quality of the PSF higher moments with two different PSF modeling algorithms (\textsc{PSFEx} and \textsc{Piff}) applied to the HSC survey dataset.  We used high-SNR star images as the true PSF, and the interpolated PSF model at the stars' position as the model PSF.  To focus on the impact of errors in the PSF higher moments, we measured the true and model PSF higher moments in a regularized coordinate system, where $e_1 = e_2 = 0$, and the second moment $\sigma$ values are the same for the model and true PSF. 
Overall, the PSF modeling quality is comparable for these methods. Our findings suggest there is value in further tuning and optimizing the PSF modeling performance for the 4\textsuperscript{th} and 6\textsubscript{th} moments for future versions of \textsc{Piff}. 

To reduce the dimensionality of the higher moment data vector and develop a basic understanding of the impact of the PSF higher moments on weak lensing, we began with  preliminary tests. 
We put an artificial Gaussian galaxy at each HSC bright star position to determine the leading PSF higher moments that affect shear measurement. Through these tests, we put 6 (5) moments into `$g_1$ group' (`$g_2$ group'), which generate additive biases on $g_1$ ($g_2$). These 11 moments also include the 7 leading moments that generate multiplicative shear biases. 

We then used the mock galaxy catalog cosmoDC2 to propagate PSF modeling errors to the weak lensing shear 2PCF. We used Gaussian Random Field to generate realizations of PSF higher moments error of the 11 aforementioned leading moments, based on their means and correlation functions measured in the HSC \textsc{PSFEx} dataset. We adopted the bulge+disc model that cosmoDC2 provides, and interpolated the shear bias for each galaxy based on their bulge size, disk size, and B/T ratio. We subdivided the cosmoDC2 galaxies into three tomographic bins to measure redshift-dependent shear biases, and found that PSF higher moment errors only generate non-zero biases in $\xi_+$. \edit{Both the multiplicative and additive biases are redshift dependent, as they all depend on the galaxy property distributions at that redshift.}

Finally, we have propagated the PSF higher moments error to systematic biases in inferred cosmological parameters using Fisher forecasting. We find that additive shear biases due to PSF higher moment errors can cause a $1\sigma$ systematic shift on key cosmological parameters, such as $\Omega_m$, $\sigma_8$ and $w_0$, at the LSST Y10 level -- implying that either PSF higher moment errors must be reduced from current levels for LSST Y10, or this effect must be explicitly modeled in the cosmological parameter analysis.  
In contrast, the multiplicative shear biases only cause cosmological parameter shifts of at most $0.1\sigma$. The forecast shows that the impact of the PSF higher moment errors on LSST Y1 is smaller than that on LSST Y10, but the effect is still not negligible even for Y1.

\edit{This work motivates several future studies:}
\begin{itemize}
\item The results of this paper imply that future surveys, including LSST and the High Latitude Survey of the Roman Space Telescope, need to design null tests to ensure that the additive shear biases due to PSF higher moment errors do not cause an unacceptable level of contamination of the weak lensing shear data vectors. Requirements on PSF higher moment modeling quality, and/or mitigation methods, are needed for these surveys to recover credible cosmological constraints from the weak lensing shear data. 
\item \edit{Modeling the PSF higher moment residuals is needed in the cosmological analyses. By cross correlating PSF higher moments residual with the estimated shear, one can measure the systematics in 2PCF associated with the PSF higher moments error, and marginalize over it in the cosmological analyses. However, the high dimensionality of this source of  systematic uncertainty remains challenging, even though this work has reduced the dimensionality by a factor of 2, encouraging future development.}
\edit{\item This work also motivates the inspection of PSF higher moment modeling quality to drive the further development of new PSF modeling algorithms. This includes inspecting whether the reconstruction, interpolation, as well as the coadding process can generate errors in the PSF higher moments.  Careful attention to this issue could greatly simplify the points mentioned above about modeling the impact of this systematic in future surveys.}
\edit{Because of the size dependence we find in both the additive and multiplicative biases, we recommend further development in redshift-dependent additive and multiplicative biases PSF systematics modeling in the cosmological analyses for the cosmic shear.}
\end{itemize}

\section*{Contributors}

TZ developed the simulation and measurement software, carried out analysis on the results, and led the writing of the manuscript. HA developed the code for the Fisher Information matrix and relevant parameter inference, and contributed writing for Section 3.5.  RM proposed the project, advised on the motivation, experimental design and analysis, and edited the manuscript. JEM advised and provided early access to HSC data processed using \textsc{Piff}.
MJ provided feedback throughout the project regarding interpretations of results, suggestions for validation tests, providing textual edition on the manuscript, and guidance on software implementation. 
AK provided ideas behind the symmetry and formalism in PSF higher moments, and feedback throughout the project. 
MAS provided feedback throughout the project, mostly in the form of questions asked regarding intermediate results and the design of the tests performed. He provided feedback and numerous suggestions on the manuscript, including changes to Figures 3 and B2, Algorithm 1, and made several minor edits.
AG provided feedback and fundamental structural suggestions to the manuscript.

\subsection*{Acknowledgments}

\refresponse{We thank the anonymous referee for their helpful feedback on this paper.} This paper has undergone internal review in the LSST Dark Energy Science Collaboration by Axel Guinot, Henk Hoekstra, and Francois Lanusse, we thank them for their constructive comments and reviews. We thank Aaron Roodman, Ares Hernandez, Xiangchong Li, Mustapha Ishak, and Douglas Clowe for the helpful comments and discussion. 

TZ and RM are supported in part by the Department of Energy grant DE-SC0010118 and in part by a grant from the Simons Foundation (Simons Investigator in Astrophysics, Award ID 620789). 

The DESC acknowledges ongoing support from the Institut National de Physique Nucl\'eaire et de Physique des Particules in France; the Science \& Technology Facilities Council in the United Kingdom; and the Department of Energy, the National Science Foundation, and the LSST Corporation in the United States.  DESC uses resources of the IN2P3 Computing Center (CC-IN2P3--Lyon/Villeurbanne - France) funded by the Centre National de la Recherche Scientifique; the National Energy Research Scientific Computing Center, a DOE Office of Science User Facility supported by the Office of Science of the U.S.\ Department of Energy under Contract No.\ DE-AC02-05CH11231; STFC DiRAC HPC Facilities, funded by UK BIS National E-infrastructure capital grants; and the UK particle physics grid, supported by the GridPP Collaboration.  This work was performed in part under DOE Contract DE-AC02-76SF00515.

Based in part on data collected at the Subaru Telescope and retrieved from the HSC data archive system, which is operated by Subaru Telescope and Astronomy Data Center at National Astronomical Observatory of Japan.

The Hyper Suprime-Cam Subaru Strategic Program (HSC-SSP) is led by the astronomical communities of Japan and Taiwan, and Princeton University.  The instrumentation and software were developed by the National Astronomical Observatory of Japan (NAOJ), the Kavli Institute for the Physics and Mathematics of the Universe (Kavli IPMU), the University of Tokyo, the High Energy Accelerator Research Organization (KEK), the Academia Sinica Institute for Astronomy and Astrophysics in Taiwan (ASIAA), and Princeton University.  The survey was made possible by funding contributed by the Ministry of Education, Culture, Sports, Science and Technology (MEXT), the Japan Society for the Promotion of Science (JSPS),  (Japan Science and Technology Agency (JST),  the Toray Science Foundation, NAOJ, Kavli IPMU, KEK, ASIAA, and Princeton University.

This paper makes use of software developed for the Vera C. Rubin Observatory. We thank the Vera C. Rubin Observatory for making their code available as free software at http://dm.lsst.org. 

The Pan-STARRS1 Surveys (PS1) have been made possible through contributions of the Institute for Astronomy, the University of Hawaii, the Pan-STARRS Project Office, the Max-Planck Society and its participating institutes, the Max Planck Institute for Astronomy, Heidelberg and the Max Planck Institute for Extraterrestrial Physics, Garching, The Johns Hopkins University, Durham University, the University of Edinburgh, Queen’s University Belfast, the Harvard-Smithsonian Center for Astrophysics, the Las Cumbres Observatory Global Telescope Network Incorporated, the National Central University of Taiwan, the Space Telescope Science Institute, the National Aeronautics and Space Administration under Grant No. NNX08AR22G issued through the Planetary Science Division of the NASA Science Mission Directorate, the National Science Foundation under Grant No. AST-1238877, the University of Maryland, and Eotvos Lorand University (ELTE) and the Los Alamos National Laboratory.

We thank the developers of \textsc{GalSim}, \textsc{ngmix}, and \textsc{TreeCorr} for making their software openly accessible. Some of the results in this paper have been derived using the \textsc{Healpy} and \textsc{HEALPix} package.

\section*{Data Availability}

The HSC-SSP data in this paper is publicly available at \url{https://hsc-release.mtk.nao.ac.jp/doc/index.php/tools-2/}. 
The COSMOS catalog is available at \url{https://zenodo.org/record/3242143#.YF2bHK9KiUk}.
The cosmoDC2 catalog is available at the LSST DESC Data Portal
\url{https://data.lsstdesc.org/}.
Simulation and analysis code is publicly available\footnote{\url{https://github.com/LSSTDESC/PSFHOME}}.

\bibliography{main}

\appendix

\section{Moment Residual Maps in HSC}
\label{ap:moment_example}

In this appendix, we show the moment residual results that are not included in the main text. In Section~\ref{ap:moment_example:fields}, we show two maps of PSF truth and residual in the HSC PDR1 data. In Section~\ref{ap:moment_example:rc2}, we show the PSF residual distribution in the HSC RC2 dataset comparing \textsc{Piff} and \textsc{PSFEx}.

\subsection{PSF Residuals by Fields}
\label{ap:moment_example:fields}

\begin{figure*}
    \centering
    \includegraphics[width=1.75\columnwidth]{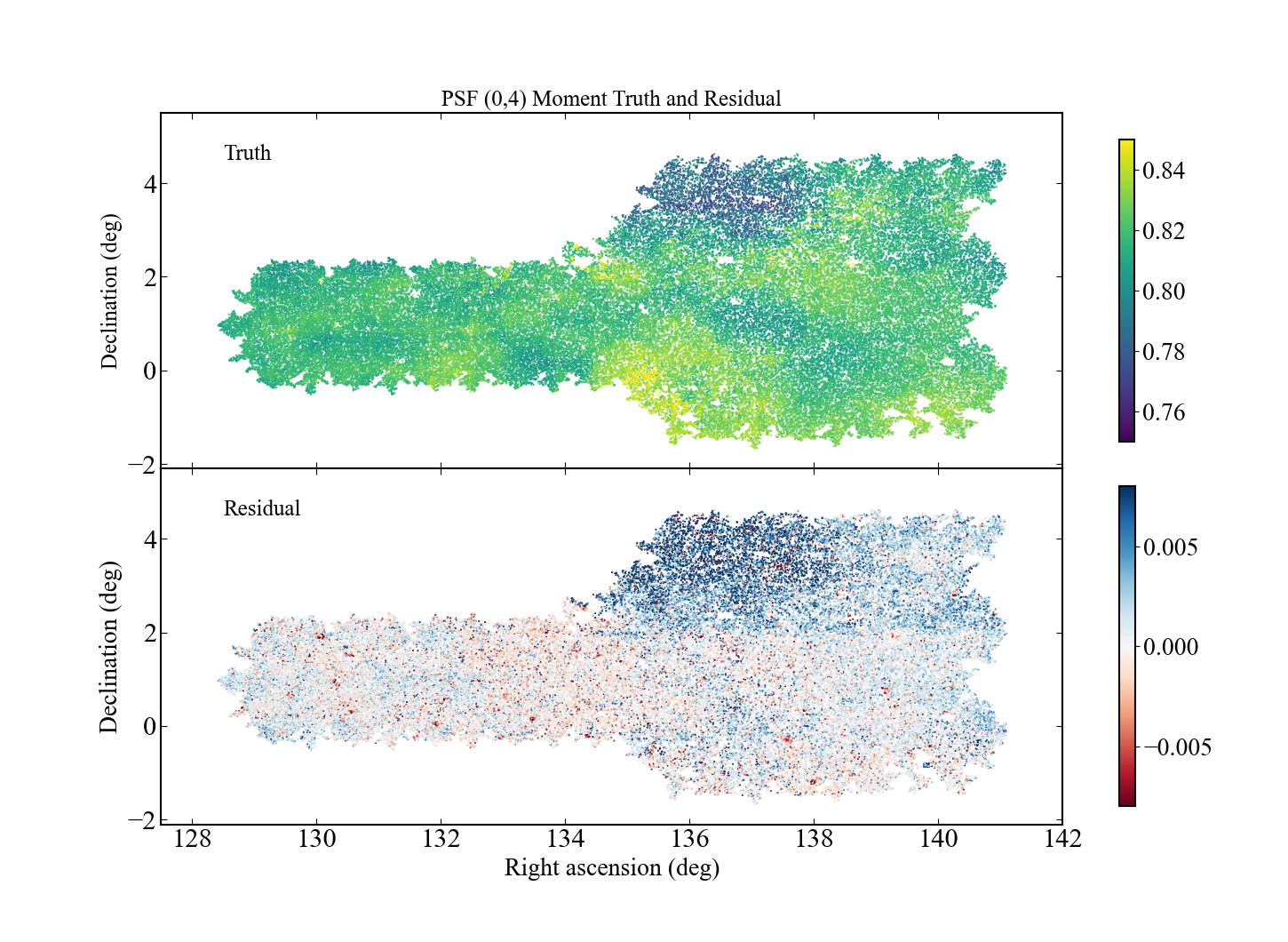}
    \includegraphics[width=2.0\columnwidth]{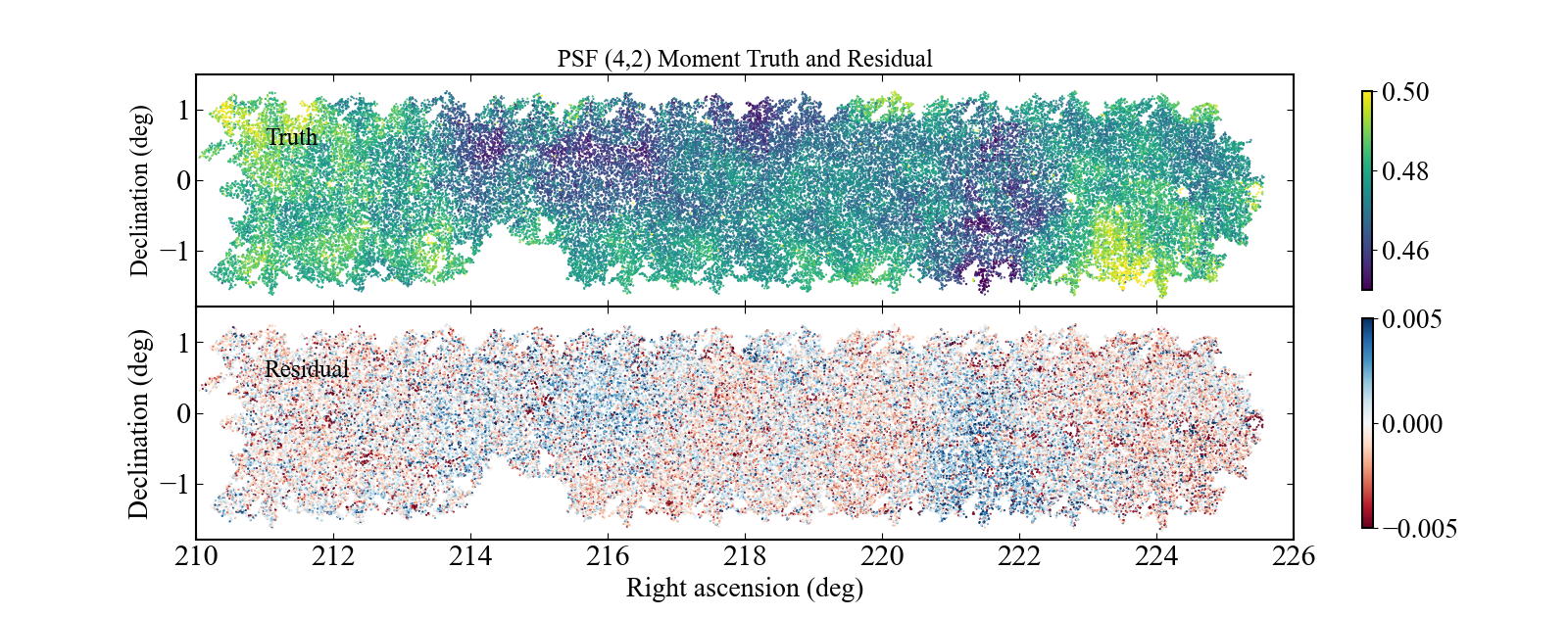}
    \caption{Two examples of the maps of PSF higher moments for the HSC PDR1 data, as modeled by \textsc{PSFEx}. For both examples, we show the true value and the moment residual $B[M_{pq}]$. The top panel shows a map of the (0,4) moment measured in the GAMA09H field, and the bottom panel shows the (4,2) moment measured in the GAMA15H field. There is coherent structure in both the true moments and their residuals, suggesting that the measurement is not noise dominated.
    \label{fig:moment_example}}
\end{figure*}

In Fig.~\ref{fig:moment_example}, 
we show two examples of the PSF moment maps that we measure in the 6 HSC fields. We show maps for the true moments and the residual $B[M_{pq}]$. We can see that the true PSF higher moments and their residuals clearly have real spatial structure, as is found in ZM21 for the radial kurtosis. The residuals are both correlated with the true moments, showing that \textsc{PSFEx} performs differently depending on the true underlying PSF, which suggests that one opportunity for improvement in future algorithms is greater performance stability.  
Comparing to figure 1 in \cite{2018PASJ...70S..25M}, we observe that the better seeing parts of the area have PSF higher moment biases that are higher than other areas, especially for the 4\textsuperscript{th} and 6\textsuperscript{th} moments. Many of these good-seeing areas are eliminated from the HSC first-year shear catalog, described in section 4 of \cite{2018PASJ...70S..25M}, due to them failing various PSF modeling tests. 
This is confirmed both visually and by the correlation matrix in Fig.~\ref{fig:correlation_matrix} for \textsc{PSFEx}. 

\subsection{Comparison between \textsc{Piff} and \textsc{PSFEx} in RC2}
\label{ap:moment_example:rc2}

\begin{figure*}
    \centering
    \includegraphics[width=1.8\columnwidth]{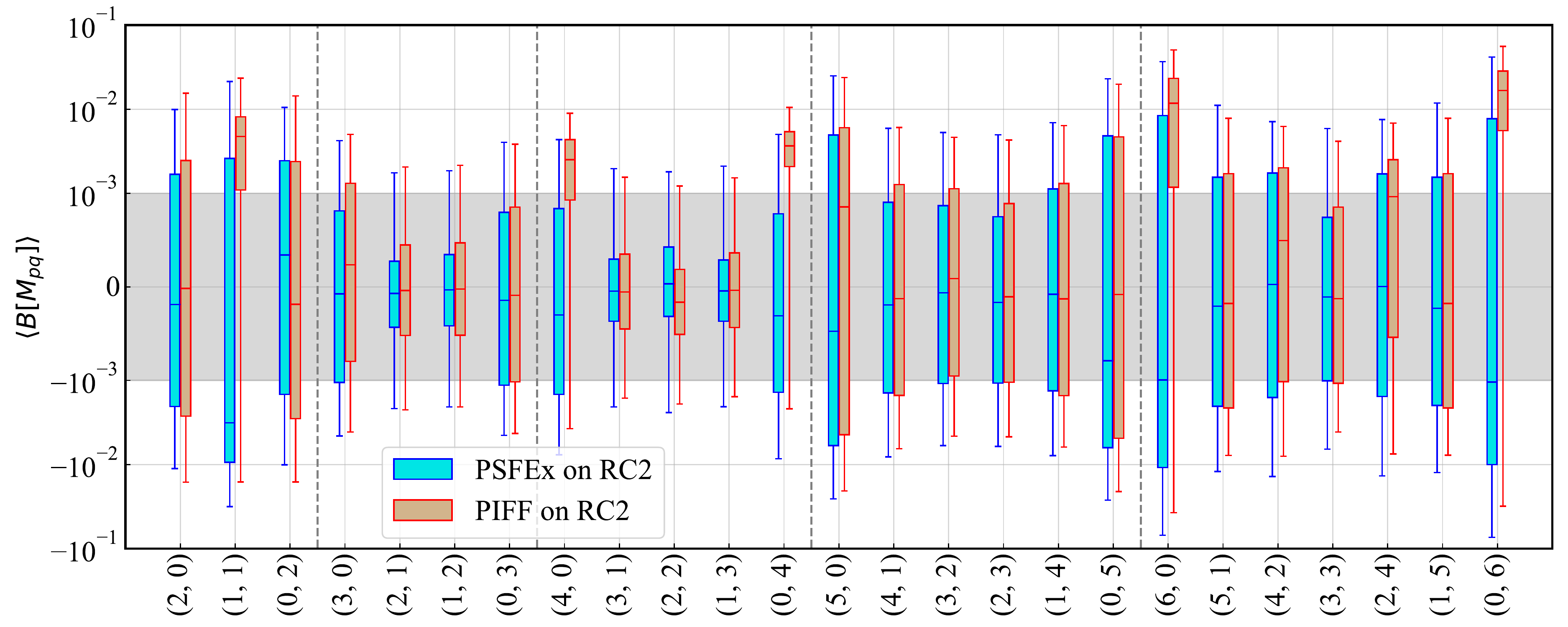}
    \caption{Box plot showing the PSF moment biases from the $2^{\text{nd}}$ to the $6^{\text{th}}$ moments, with the whiskers showing the $2\sigma$ range (from 3rd to 97th percentile), the boxes showing the interquartile range, and the bars showing the median. The \textsc{PSFEx} and \textsc{Piff} results, both runned on the RC2 dataset described in Section~\ref{sec:data:piff}, are shown side-by-side. The y-axis is symmetrical log-scaled, with the linear region shown in grey. 
    }
    \label{fig:rc2_comparison}
\end{figure*}

\edit{In Figure~\ref{fig:rc2_comparison}, we show an apples-to-apples comparison between \textsc{Piff} and \textsc{PSFEx} on the RC2 dataset. This is in addition to the comparison made in Section~\ref{sec:data:hsc_measure}. Due to their settings, \textsc{PSFEx} and \textsc{Piff} have opposite signs in many key moments for the weak lensing systematics. However, \textsc{PSFEx} shows lower moment residuals compared to \textsc{Piff}. This further motivates the development and optimization of \textsc{Piff}, which when properly tuned should improve in performance. }

\section{Shapelet-Moment Relation}
\label{ap:shapelet_moment}

\begin{figure*}
    \centering
    \includegraphics[width=2.0\columnwidth]{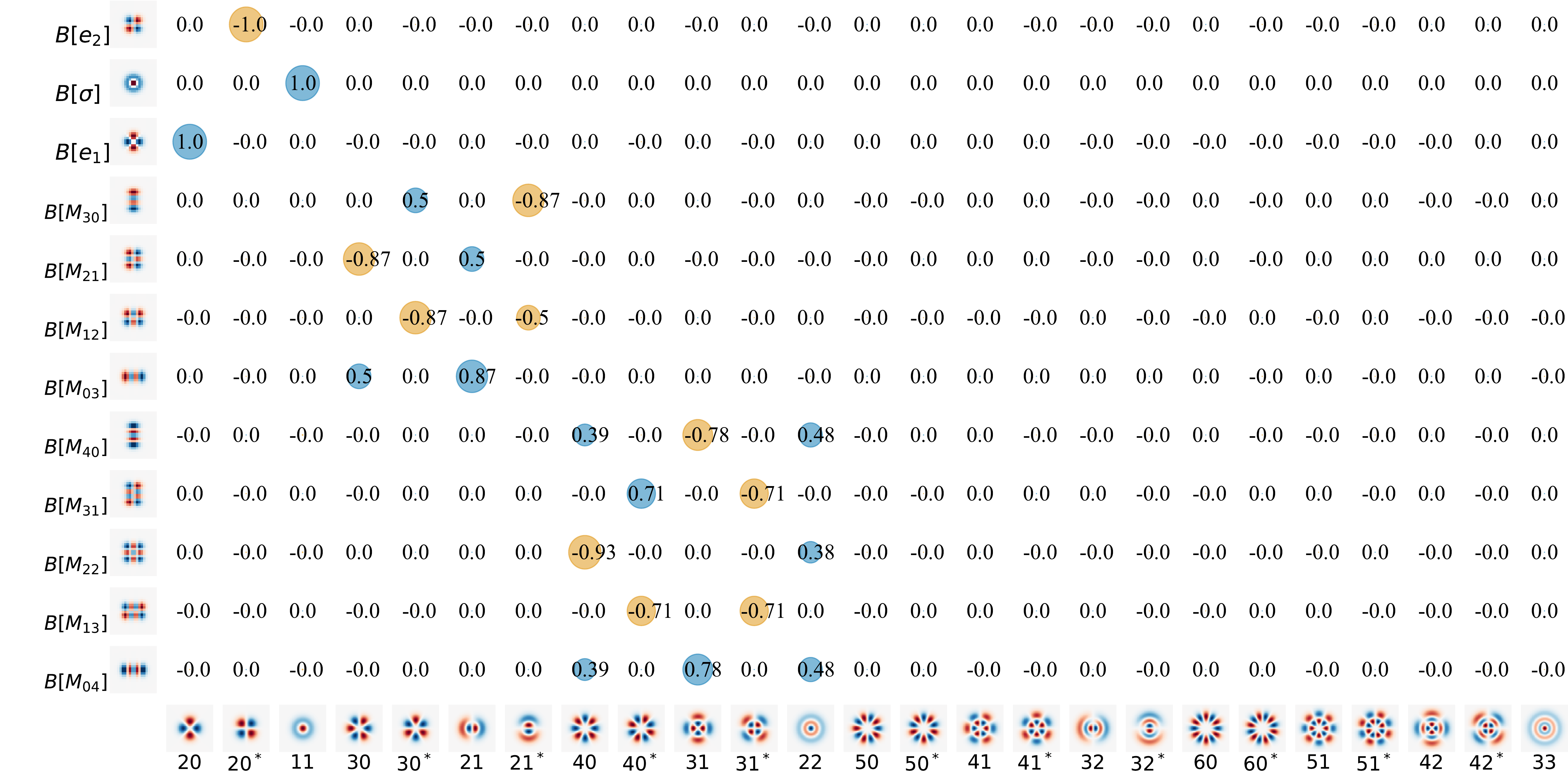}
    \caption{The Jacobian of PSF moments with respect to the shapelet coefficients, 
    $T_{pq,jk} = \frac{\partial M_{pq}}{\partial b_{jk}}$ for a Kolmogorov PSF. We show the first 12 PSF moments starting from the second moments (i.e., 3 second moments, 4 third moments, and 5 fourth moments) 
    and the first 25 shapelet components starting from $n = 2$ for the shapelets decomposition.  The numbers that overlap the circle are the values of $T_{pq,jk}$, with each row normalized by the $L^2$-norm $\sqrt{\sum_{jk} T_{pq,jk}^2}=1$. The sizes of the circles reflect the magnitude of the entry, and colors reflect the sign (blue for positive and yellow for negative).   The column on the left shows the postage stamp images of the difference in PSF with only one moment being changed. The bottom row shows the first 25 shapelet bases, as the bases for the moment modification. We rank the shapelet coefficients by increasing the order $n$. For each $n$, we start with the real part of $j = n$, then its imaginary part, and decrease $j$ until $j=k$ or $j=k+1$. The labels on the shapelet basis functions should be interpreted as follows: $jk$ is equivalent to Re$[\psi_{jk}]$, $jk^*$ is equivalent to Im$[\psi_{jk}]$.  We can see that the Jacobian matrix is very close to being a  block-diagonal matrix, which means that the PSF higher moments are linear combinations of the shapelet components with the same order $n$.
    \label{fig:Tij}
    }
\end{figure*}

In Section~\ref{sec:simulation:mom_method}, we explained that the key to changing the PSF moments through a shapelet decomposition is to compute the Jacobian matrix using Eq.~\eqref{eq:jaco_cal}.
In Fig.~\ref{fig:Tij}, we show an example of the Jacobian matrix $T$ for PSF higher moment errors $B[M_{pq}]$ with respect to the shapelet modes $b_{jk}$  
for a Kolmogorov PSF. There is a block diagonal structure that shows the PSF second moments depend on the $\text{4}^{\text{th}}$ to $\text{5}^{\text{th}}$ shapelet modes. The PSF third moments 
depend on the $\text{6}^{\text{th}}$ to $\text{9}^{\text{th}}$ shapelet modes. The PSF fourth moments
depend not only on the $\text{10}^{\text{th}}$ to $\text{14}^{\text{th}}$ shapelet modes, but also on the shapelet modes that determine the second moments. This means that the n\textsuperscript{th} PSF higher moments basis can be approximately decomposed into shapelets components with the same order $n$. 

With the Jacobian matrix shown in Fig.~\ref{fig:Tij} and the Algorithm~\ref{alg:change_moment}, we modify the individual PSF higher moments with moment error threshold $||\mathbf{\Delta M}||_2 = 10^{-6}$. 
This precision is sufficient for exploring systematic shear biases associated with errors in the PSF higher moments.

\section{Symmetry in the response to PSF higher moments}
\label{sec:app:symmetry}

In Fig.~\ref{fig:shear_response}, it is clear that the shear response to the PSF higher moments exhibits symmetries among the different higher moments. In this section, we explore and explain this symmetry. We start by proposing four lemmas, and derive the symmetry of the shear response based on these four lemmas.

Lemma 1: For any two PSF modeling residual basis functions $B[M_{pq}](x,y)$ and $B[M_{uv}](x,y)$ such that $B[M_{pq}](x,y)$ can be obtained by rotating $B[M_{uv}](x,y)$ by $\pm90$~\deg, the corresponding shear biases $\Delta \hat{g}(B[M_{pq}])$ and $\Delta \hat{g}(B[M_{uv}])$ satisfy the following constraint:
\begin{equation}
    \Delta \hat{g}(B[M_{pq}]) = - \Delta \hat{g}(B[M_{uv}])
\end{equation}
Note that this Lemma is also the basis for the fact that a 90~\deg rotated galaxy pair has an average shape of $0$, a fact that we use in the single galaxy simulations. We also stress that the $B[M_{pq}](x,y)$ in this section is a functional basis of the higher moments error (see Fig.~\ref{fig:moment_basis} for examples), different from the moment biases $B[M_{pq}]$ elsewhere.

Lemma 2: The PSF modeling residual function $B[M_{pq}](x,y)$ has the form
\begin{align}
    B[M_{pq}](x,y) &= B[M_{pq}](-x,y) \; \; \; \;& \text{if p is even}\\
    B[M_{pq}](x,y) &= -B[M_{pq}](-x,y) \; \; \; & \text{if p is odd.}
\end{align}
This is due to the symmetry (asymmetry) in the even (odd) functions used in the moment measurement. 

Lemma 3: Similar to Lemma 2, the PSF modeling residual function $B[M_{pq}](x,y)$ 
has the form:
\begin{align}
    M_{pq}(x,y) &= M_{pq}(x,-y) \; \; \; \;& \text{if q is even}\\
    M_{pq}(x,y) &= -M_{pq}(x,-y) \; \; \; & \text{if q is odd.}
\end{align}

Lemma 4: For any $p$ and $q$, the PSF modeling residual function satisfies the relationship 
\begin{equation}
    B[M_{qp}](x,y) = B[M_{pq}](y,x). 
\end{equation}
This is easily proved by substituting $y$ for $x$ and vice versa.

With these lemmas, we can identify  symmetry relationships between different moments, as long as the moment responses are rotations of each other. But first, we must define the rotation operators $\mathcal{R}$ and $\mathcal{R}'$,
\begin{itemize}
    \item $\mathcal{R}(B[M_{pq}](x,y)) = B[M_{pq}](y,-x)$ -- a function that rotates the moment response by $90$~deg clockwise.  
    \item $\mathcal{R}'(B[M_{pq}](x,y)) = B[M_{pq}](-y,x)$ -- a function that rotates the moment response by $90$~\deg counter-clockwise.
\end{itemize}

 Fig.~\ref{fig:shear_response} shows that the symmetry between results for different moments depends on the parity of the moment index $p$ and $q$. Therefore, we consider four scenarios with different parities.

Case 1: If both $p$ and $q$ are even:
\begin{align}
    B[M_{qp}](x,y) &= B[M_{pq}](y,x) \\\nonumber &= B[M_{pq}](y,-x) = \mathcal{R}(B[M_{pq}](x,y))
\end{align}
The first two steps use Lemma 4 and Lemma 3, respectively. The last step relies on the definition of  $\mathcal{R}$. Using Lemma 1 on the very left-hand-side (LHS) and very right-hand-size (RHS) of this equation, we infer that $\Delta\hat{g}(B[M_{pq}]) = -\Delta\hat{g}(B[M_{qp}])$, which is consistent with the results for $(p,q)=(0,4)$ in Fig.~\ref{fig:shear_response}. This case also implies that for even values of $p=q$, $\Delta\hat{g}$ must be 0, as is seen for the $(p,q)=(2,2)$ case in Fig.~\ref{fig:shear_response}.

Case 2: If $p$ is even and $q$ is odd:
\begin{align}
    B[M_{qp}](x,y) &= B[M_{pq}](y,x)\\\nonumber &= B[M_{pq}](-y,x) = \mathcal{R}'(B[M_{pq}](x,y)).
\end{align}
The first two steps utilize Lemma 4 and Lemma 2, respectively. The last step relies on the definition of $\mathcal{R}'$. With Lemma 1 applied to the very LHS and RHS, we infer that $\Delta\hat{g}B[(M_{pq}]) = -\Delta\hat{g}(B[M_{qp}])$ for this case. This finding is consistent with the results for $(p,q)=(0,3)$ and $(2,1)$ in Fig.~\ref{fig:shear_response}.

Case 3: If $p$ is odd and $q$ is even, the only difference from Case 2 is to flip $x$ instead of $y$ in the second step:
\begin{align}
    B[M_{qp}](x,y) &= B[M_{pq}](y,x)\\\nonumber &= B[M_{pq}](y,-x) = \mathcal{R}(B[M_{pq}](x,y)).
\end{align}
The first two steps utilize Lemma 4 and Lemma 3. The last step relies on the definition of $\mathcal{R}$. With Lemma 1 applied to the very LHS and RHS, we infer that $\Delta\hat{g}(B[M_{pq}]) = -\Delta\hat{g}(B[M_{qp}])$. 

Case 4: If both $p$ and $q$ are odd:
\begin{align}
   B[M_{qp}](x,y) &= B[M_{pq}](y,x)\\\nonumber &= -B[M_{pq}](y,-x) = - \mathcal{R}(B[M_{pq}](x,y)).
\end{align}
The first two steps use Lemma 4 and Lemma 3. The last step relies on the definition of $\mathcal{R}$. Applying Lemma 1 to the very LHS and RHS, we infer that $\Delta\hat{g}(B[M_{pq}]) = \Delta\hat{g}(B[M_{qp}])$. This finding is consistent with the results for $(p,q)=(1,3)$ in Fig.~\ref{fig:shear_response}. 

In conclusion, only when both $p$ and $q$ are odd will we get $\Delta\hat{g}(B[M_{pq}]) = \Delta\hat{g}(B[M_{qp}])$. Otherwise, $\Delta\hat{g}(B[M_{pq}]) = -\Delta\hat{g}(B[M_{qp}])$, implying that pairings with even values of $p=q$ produce zero shear bias.  As described above, these symmetry patterns are displayed in Fig.~\ref{fig:shear_response}. While not shown in the plot, we have explicitly confirmed that the above conclusions apply to the 5\textsuperscript{th} and 6\textsuperscript{th} moments as well, and they should hold beyond that as well.

\section{Generating the Gaussian Random Fields}
\label{ap:grf}

\begin{figure*}
    \centering
    \includegraphics[width=1.3\columnwidth]{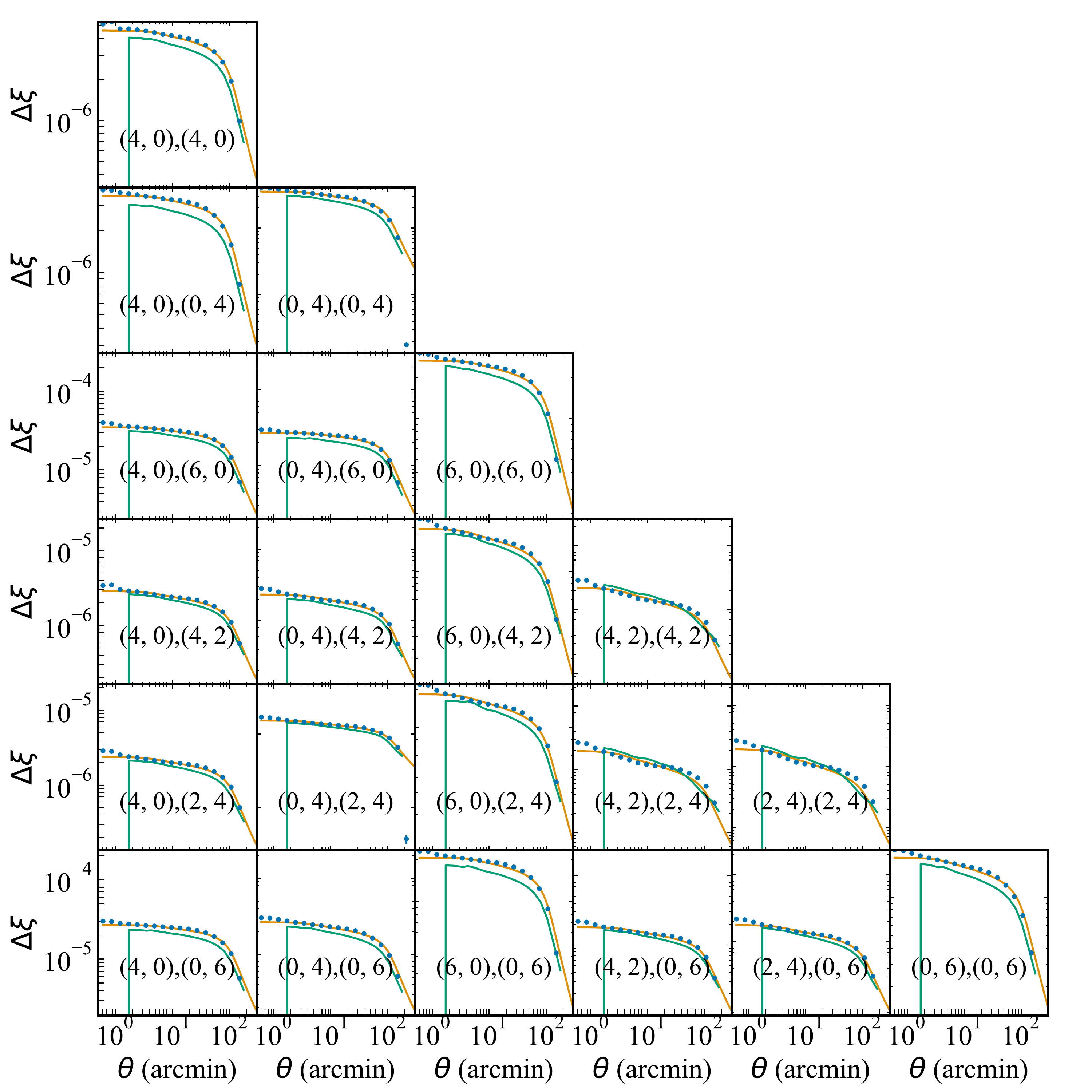}
    \includegraphics[width=1.3\columnwidth]{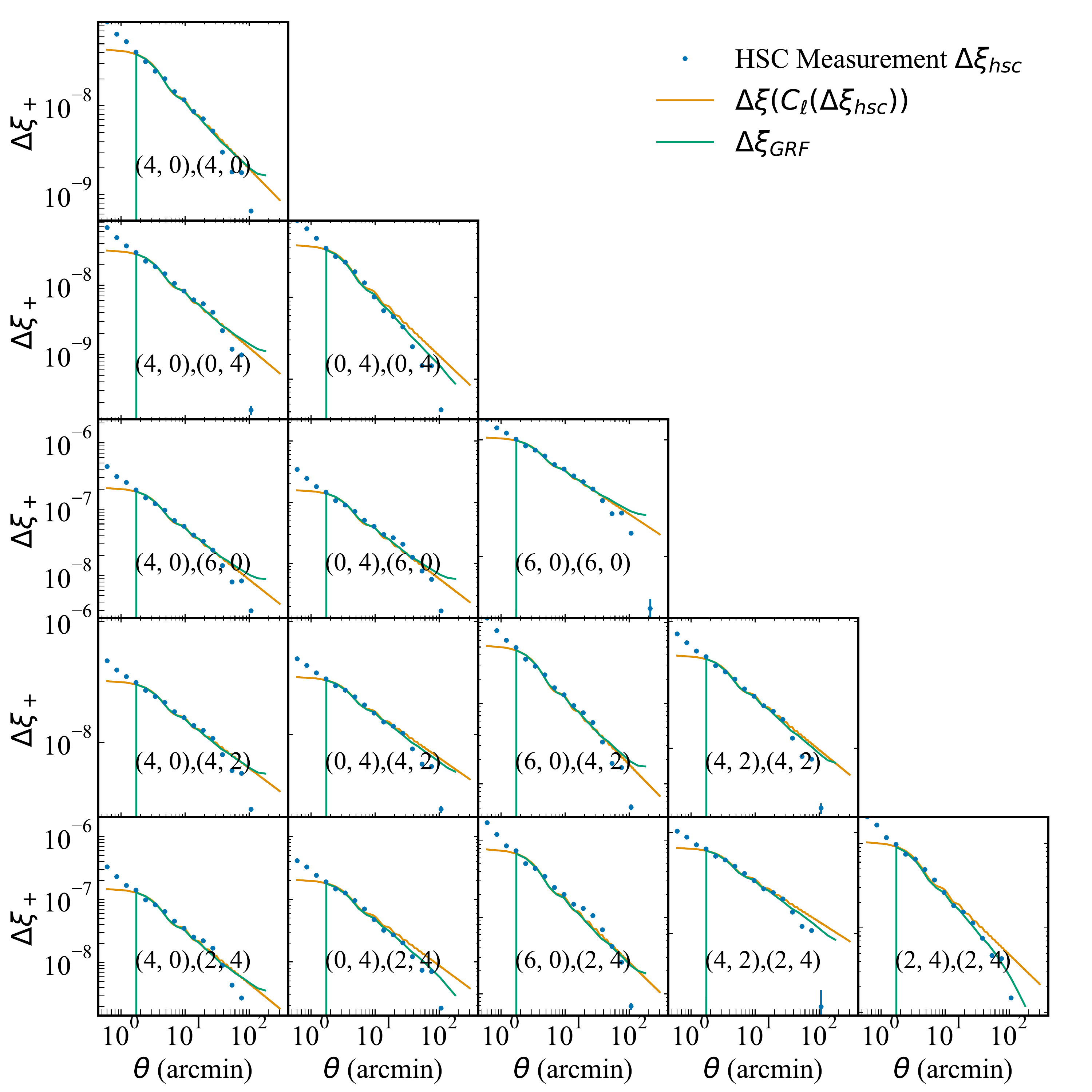}
    \caption{The original correlation function of the PSF higher moment residual fields measured from all six HSC PDR1 fields combined  
    (blue dots), the ``round-trip transformation'' of the original correlation functions 
    (orange curve), and the PSF moments residual correlation functions 
    of the generated GRF. The upper panel shows the `$g_1$ group' moments, and the lower panel shows the `$g_2$ group' moments that we defined in Section~\ref{sec:analyses:reduction}. 
    The 2PCFs of the GRFs generally match those in the HSC data, except at angular scales $\le 1.7$ arcmin, corresponding to the scale of the \textsc{HEALPix} grid. 
    \label{fig:generated_map}}
\end{figure*}

In Fig.~\ref{fig:generated_map}, 
we show the correlation functions of the PSF higher moment residual maps, described in Eq.~\eqref{eq:moment_2pcf}, of the two groups of moments that determine the additive shear biases, defined in Section~\ref{sec:analyses:reduction}. These two groups of PSF higher moments are later used to propagate PSF higher moment error to the cosmoDC2 galaxies in Section~\ref{sec:analyses:mock}. 
The blue dots are the measurements based on the HSC bright stars and \textsc{PSFEx} in Section~\ref{sec:data:hsc_measure}. 

We have devised empirical fitting formulae to describe the measurements of the correlation functions shown as blue dots.  We fit the `$g_1$ group' correlation functions with a power-arctan function,
\begin{equation}
    \xi_{\text{fit}}(\theta) = a \theta^{-b} \left[\frac{1}{2}- \frac{ \text{tan}^{-1}(\theta - \theta_\text{cutoff})}{\pi} \right].
\end{equation}
This model is chosen because the correlation function takes the form of a power law on small scales (or a linear function when plotted on a log-log plot), and then rapidly drops to zero. The part in the parenthesis is designed to produce the rapid drop to zero on scales beyond  $\theta_\text{cutoff} = 1.7$~\deg, a scale that is comparable to the size of the Subaru FOV.  For the `$g_2$ group' correlation functions, we use a power law, 
\begin{equation}
    \xi_{\text{fit}}(\theta) = a \theta^{-b},
\end{equation}
as correlation functions in the `$g_2$ group' moments are visually consistent with a power law.

Since the cosmoDC2 catalog has an area larger than any field in the HSC data we measure, we need to generate artificial PSF moment residual maps to cover the cosmoDC2 area.  We convert the above fitting functions to angular power spectra by carrying out a Hankel transform, using \textsc{SkyLens}\footnote{\url{https://github.com/sukhdeep2/Skylens\_public/tree/imaster\_paper/}} \citep{2021MNRAS.508.1632S}, 
and use the power spectra to generate artificial Gaussian Random Fields (GRF) using \textsc{Healpy} \citep{Zonca2019,2005ApJ...622..759G}. 
To ensure the integration is stable and bug-free, we do a round-trip transformation test, where we transform the power spectrum back to real space, shown in the orange curves in Fig.~\ref{fig:generated_map}. The round-trip transformations match the original data well in all cases, as a validation of the fitting function and the numerical accuracy of these transformations. The GRF is generated on the \textsc{HEALPix} sphere with $N_{\text{side}} = 2048$, with a pixel size $\sim 1.7$ arcmin. We measure the correlation function of the GRF, shown as the green curves in Fig.~\ref{fig:generated_map}. Except for the angular bins that are below the resolution of the \textsc{HEALPix} grid,  
the GRF is shown to match the original field well in terms of the two-point statistics.

\label{LastPage}

\end{document}